\documentclass[11pt,a4paper]{article}

\usepackage[utf8]{inputenc}
\def\papertitle{\bf Final-state rescattering mechanism in \deleted{of} bottom-baryon decays} 
\usepackage{authblk}
\title{\papertitle}
\author[1]{Zhu-Ding Duan, \footnote{Email: 32246002@mail.imu.edu.cn}}
\author[2]{Jian-Peng Wang\footnote{Email: wangjp20@lzu.edu.cn, corresponding author}}
\author[1]{Run-Hui Li\footnote{Email: lirh@imu.edu.cn, corresponding author}}
\author[3,4]{Cai-Dian L\"u\footnote{Email: lucd@ihep.ac.cn, corresponding author}}
\author[2]{Fu-Sheng Yu\footnote{Email: yufsh@lzu.edu.cn, corresponding author}}

\affil[1]{Center for Quantum Physics and Technologies, School of Physical Science and Technology, Inner Mongolia University, Hohhot 010021, China}
\affil[2]{Frontiers Science Center for Rare Isotopes, and School of Nuclear Science and Technology, Lanzhou University, Lanzhou 730000, China}
\affil[3]{Institute of High Energy Physics, CAS,
P.O. Box 918(4) Beijing 100049, China}
\affil[4]{School of Physical Sciences, University of Chinese Academy of Sciences, Beijing 100049, China}
\usepackage[final]{changes}
\usepackage[top=1in, bottom=1.25in, left=1in, right=1in]{geometry}
\usepackage{microtype}
\usepackage{lineno}
\usepackage{xspace}
\usepackage{booktabs}
\usepackage{comment}
\usepackage{slashed}

\usepackage{caption} 

\usepackage{graphicx}  

\usepackage{color}
\usepackage{colortbl}
\DeclareGraphicsExtensions{.pdf,.PDF,png,.PNG}

\usepackage{amsmath} 
\usepackage{amsfonts,amssymb,amsthm,bm,braket,yhmath}
\usepackage{upgreek} 

\usepackage{hyperref}
\usepackage{hyperxmp}
\usepackage[figuresright]{rotating}
\usepackage{hypcap} 
\hypersetup{hidelinks} 
\usepackage{txfonts}
\usepackage{amssymb}
\usepackage{cases}
\usepackage{algorithm}
\usepackage{algpseudocode}
\usepackage{mathrsfs}
\usepackage{color}
\usepackage{amscd,amssymb,amsthm,amsmath,bm,graphicx,psfrag}
\usepackage{epsfig,mathrsfs}
\usepackage[bf,SL,BF]{subfigure}
\usepackage{amsmath}
\usepackage{pict2e,keyval,fp}
\usepackage[bf,SL,BF]{subfigure}

\theoremstyle{definition}

\numberwithin{equation}{section}

\def\kaon   {{\ensuremath{K}}\xspace}
\def\Kbar    {{\kern 0.2em\overline{\kern -0.2em \kaon}{}}\xspace}

\def\KorKbar    {\kern 0.18em\optbar{\kern -0.18em K}{}\xspace}

\def\Dbar    {{\kern 0.2em\overline{\kern -0.2em D}{}}\xspace}

\def\DorDbar    {\kern 0.18em\optbar{\kern -0.18em D}{}\xspace}

\def\Lbar        {{\ensuremath{\kern 0.1em\overline{\kern -0.1em\Lambda}}}\xspace}
\def\LorLbar    {\kern 0.18em\optbar{\kern -0.18em \PLambda}{}\xspace}

\newcommand{\tev}{\ensuremath{\mathrm{\,Te\kern -0.1em V}}\xspace}
\newcommand{\gev}{\ensuremath{\mathrm{\,Ge\kern -0.1em V}}\xspace}
\newcommand{\mev}{\ensuremath{\mathrm{\,Me\kern -0.1em V}}\xspace}
\newcommand{\kev}{\ensuremath{\mathrm{\,ke\kern -0.1em V}}\xspace}
\newcommand{\ev}{\ensuremath{\mathrm{\,e\kern -0.1em V}}\xspace}
\newcommand{\gevc}{\ensuremath{{\mathrm{\,Ge\kern -0.1em V\!/}c}}\xspace}
\newcommand{\mevc}{\ensuremath{{\mathrm{\,Me\kern -0.1em V\!/}c}}\xspace}
\newcommand{\gevcc}{\ensuremath{{\mathrm{\,Ge\kern -0.1em V\!/}c^2}}\xspace}
\newcommand{\gevgevcccc}{\ensuremath{{\mathrm{\,Ge\kern -0.1em V^2\!/}c^4}}\xspace}
\newcommand{\mevcc}{\ensuremath{{\mathrm{\,Me\kern -0.1em V\!/}c^2}}\xspace}

\usepackage{cite} 
\usepackage{mciteplus}

\begin{document}
\maketitle
\begin{abstract}

  The first observation of $CP$ violation in baryon decays was recently reported by the LHCb collaboration in $\Lambda_b^0\to pK^-\pi^+\pi^-$ with $A_{CP} = (2.45 \pm 0.46 \pm 0.10)\%$, which inspires the study on baryon non-leptonic decays. In this work, we perform the first calculation of five exclusive non-leptonic decays, $\Lambda^{0}_{b}\to p\pi^{-}, p K^{-},p\rho^{-},pK^{*-}$ and $\Lambda\phi$, within the re-scattering approach. The triangle diagrams at hadron level are calculated in form of loop integrations. It leads to the generation of strong phases, which is essential to the calculation of $CP$ asymmetries. We present numerical results for branching ratios, direct and partial-wave $CP$ asymmetries, decay asymmetry parameters and their associated $CP$ asymmetries. Our results are consistent with the current LHCb experimental data, which indicates the validity and potential of our approach in studying the $CP$ asymmetries of $b$-baryon decays. Most of our results are expected to be tested in future experiments.

\newpage

\end{abstract}
\tableofcontents
\section{Introduction}

The $CP$ violations ($CPV$s) have been well-established in $K$, $B$, and $D$ meson decays over the past decades. With the data accumulation of bottom baryons at the Large Hadron Collider (LHC), some charmless non-leptonic decays of $\Lambda^{0}_{b}$ have been investigated in search for $CPV$s in baryon decays, which covers the two-body decays \cite{LHCb:2018fly, LHCb:2024iis}, three-body decays \cite{LHCb:2014yin, LHCb:2016rja, LHCb:2024yzj}, and four-body decays \cite{LHCb:2025ray, LHCb:2019oke, LHCb:2018fpt, LHCb:2016yco}. Recently, the LHCb collaboration reported the first observation of $CPV$ in baryon decays with $A_{CP} = (2.45 \pm 0.46 \pm 0.10)\%$ in $\Lambda_b^0 \to pK^-\pi^+\pi^-$ \cite{LHCb:2025ray}, which inspires the study on $b$-baryon decays. \deleted{Besides its special meaning to $CPV$ research, study on b-baryon decays can be complementary research on the flavor anomalies in $B$ meson decays such as $R(K^{(*)})$ and $R(D^{(*)})$\cite{Belle:2016dyj,Belle:2016fev,LHCb:2021trn}.} Compared to meson decays, the baryon decays exhibit richer helicity structures in their amplitudes, which facilitates the construction of more observables \cite{Gronau:2015gha,Geng:2021sxe,Ajaltouni:2004zu,Wang:2022fih,Rui:2022jff,Durieux:2016nqr,LHCb:2016hwg}. The baryon decays also manifest interesting phenomena~\cite{Yu:2024cjd}, which provides effective insights for QCD studies and precision tests of the Standard Model.

On the theoretical side, $b$-baryon decays have been studied for a long time. Most of the papers are focused on the semi-leptonic decays \cite{Detmold:2015aaa,Wang:2009hra,Khodjamirian:2011jp,Huang:2022lfr,Han:2022srw,Lu:2009cm,Shih:1998pb,Feldmann:2011xf}. However, the predictions of non-leptonic decays of $b$-baryons remains a highly challenging task. It is because that baryons have more valence quarks, introducing more complicated QCD dynamics. Additionally, non-factorizable and charm penguin contributions cannot be neglected when calculating $CP$ asymmetries\cite{Cheng:2004ru} \deleted{\cite{Leibovich:2003tw,Cheng:1996cs,
Cheng:1992gv,Lu:2000em,Keum:2000wi,Beneke:2003zv}}. Several methods have been developed to calculate $b$-baryon decays such as QCDF\cite{Zhu:2018jet}, $SU(3)$ flavor symmetry\cite{Roy:2019cky,He:2015fsa}, generalized factorization \cite{Geng:2016gul,
Hsiao:2017tif}, PQCD \cite{Lu:2009cm,Yu:2024cjd} and others. In our previous work \cite{Jia:2024pyb}, we developed a framework that described the final state interactions (FSIs) towards hadronic loop, and applied it to charm baryon decays successfully.

 The approach of FSIs has several advantages in the study of baryon decays. Firstly, it provides a systematic approach to calculate the non-factorizable and charm penguin contributions \cite{Cheng:2004ru}. Based on the experience from $B$ meson decay studies, these contributions are likely crucial for investigating $CP$ asymmetries\cite{Cheng:2004ru}. Secondly, it provides a natural picture for strong phases arising from hadronic scatterings, which are essential for direct $CPV$ in hadron decays. It is different from that of perturbative loop contributions in QCD factorization, known as BSS mechanism \cite{Bander:1979px}. Thirdly, it also provides a framework for calculating multi-body decays of baryons by incorporating quasi-two-body intermediate sub-processes.

\deleted{The principal motivation of studying the weak decays of bottom hadrons, in the context of heavy flavor physics, is that it can provide an ideal platform to extract the quark-mixing Cabibbo-Kobayashi-Maskawa (CKM) matrix elements, explore various aspects of methods and models 
 on strong dynamics developed in the past decades. Furthermore, it can be used to test the Standard Model (SM) and search for signals of new physics from an aspect of accuracy. $CP$ violation ($CPV$), implemented through an unique irreducible phase in the CKM matrix of SM\cite{Cabibbo:1963yz,Kobayashi:1973fv}, is essential when one wants to understand the dynamical origin of baryon anti-baryon asymmetry observed in the universe \cite{Sakharov:1967dj}, and also sensitive to the strong phases whose theoretical values are usually model-dependent. Therefore, how to \textcolor{red}{making} reliable calculations on strong interaction dynamics is very important. }

\deleted{
The investigations on b-baryon weak decays will reveal rich dynamical information of baryon exclusive decays, some of which may be quite different from those of mesonic systems. The theoretical frameworks of baryon decays are expected to be more complicated than those of meson decays since each baryon has three valence quarks. Additionally, the non-zero spin of the baryons brings in opportunities to analyze the helicity structures of effective Hamiltonian and some observables, and construct more triple products observables as well\cite{Gronau:2015gha,Geng:2021sxe,Ajaltouni:2004zu,Wang:2022fih,Rui:2022jff,Durieux:2016nqr,LHCb:2016hwg}. As an distinctive example, the possible cancellation between $S$ and $P$ wave $CP$ asymmetries in $\Lambda^{0}_{b}\to p\pi^{-}$ and $pK^{-}$ which is firstly discovered in Ref.~\cite{Yu:2024cjd} and also confirmed in this work. Meanwhile, there are some \textcolor{red}{anomalies} \deleted{tensions} in $B$ decays \cite{Belle:2016dyj,Belle:2009zue,Belle:2016fev,LHCb:2014vgu,LHCb:2017avl,LHCb:2015gmp,LHCb:2015svh,LHCb:2021trn}, which might indicate the existence of new physics. It is necessary and meaningful to check these phenomena in b-baryon systems. }

\deleted{More and more bottom baryons are generated at the Large Hadron Collider (LHC), and several $\Lambda^{0}_{b}$ charmless non-leptonic decays are observed \cite{LHCb:2016hwg,LHCb:2014yin,LHCb:2018fly,LHCb:2015kmm,LHCb:2016rja,
LHCb:2017gjj}.  It is anticipated that more precise investigations, especially the first confirmation of $CP$ violation in the baryon decays, can be performed. Analogous to $B\to \pi\pi$ and $K\pi$ decays, whose $CPV$s are caused by the interference of amplitudes with different weak and strong phases, $\Lambda^{0}_{b}\to p\pi^{-}$ and $pK^{-}$ are highly promised to search for baryon $CPV$, and the associated experimental observation has been performed within uncertainty of the order $1\%$ \cite{LHCb:2018fly}. Nevertheless, it is not enough to lead to a definite confirmation of baryon CPV, which is out of expectation from trivial extension of $B$ meson experience. 
In recent years, in order to observe the $CPV$ in baryon system, experimentalists checked a lot of  exclusive $b-$ baryon multibody decays, such as $\Lambda^{0}_{b}\to pK\pi\pi,pKKK,p\pi\pi\pi$... An evidence with the confidence level of $3.3\sigma$ gained in $\Lambda^{0}_{b}\to p\pi\pi\pi$ deacy inspires largely  the interests on these decays in theoretical studies \cite{LHCb:2019oke,LHCb:2018fpt,LHCb:2016yco,LHCb:2017slr}. Many  phenomenological methods are proposed, such as complete angular distribution analysis with polarization\cite{Geng:2021sxe}, $T-$odd triple product correlations\cite{Wang:2022fih}, and possible observation of $CPV$ from interference among different intermediate resonances appearing in some decays \cite{Durieux:2016nqr
,Zhao:2024ren,Zhang:2022iye,Zhang:2022emj}. However, the absence of definitive dynamical predictions brings in a lot of controversies for these \textcolor{red}{theoretical methods}\deleted{phenomenological opinions}. The discovery of large $CP$ asymmetry in $B\to \pi\pi\pi$ decay \cite{LHCb:2019jta,
LHCb:2022fpg,AlvarengaNogueira:2015wpj,Cheng:2020ipp,Garrote:2022uub,
Bediaga:2022sxw} indicates that one can investigate \textcolor{red}{$\Lambda^{0}_{b}\to p\pi\pi^{-},p\pi K^{-},p2\pi\pi^{-},p2\pi K^{-}$} decays by employing the experimental data of low energy $N\pi$ scatterings to reduce theoretical uncertainties \cite{Wang:2024oyi}. Unfortunately, because of the lack of experimental data, a theoretical approach based on dynamical calculation is still required in order to make the $CPV$ predictions for multibody $\Lambda^{0}_{b}$ decays.}

\deleted{On the theoretical side, many advances have been made in both semi-leptonic and non-leptonic decays. The main task in calculation of the semi-leptonic decays is the determination of heavy to light form factors, which are non-perturbative and extensively studied with various QCD approaches and quark models such as Lattice QCD \cite{Detmold:2015aaa}, QCD sum rules \cite{Dosch:1997zx,Huang:1998rq,MarquesdeCarvalho:1999bqs}, light cone QCD sum rules (LCSR) \cite{Huang:2004vf,Wang:2009hra,Azizi:2009wn,
Khodjamirian:2011jp,Huang:2022lfr}, perturbative QCD approach (PQCD) \cite{Han:2022srw,
Lu:2009cm,Shih:1998pb}, soft-collinear effective theory (SCET) \cite{Feldmann:2011xf}, light front quark model (LFQM) \cite{Zhu:2018jet}, non-relativistic quark model \cite{Cheng:1995fe}, and MIT bag model \cite{Geng:2020ofy}. It is interesting that high twist contributions are dominating in $\Lambda^{0}_{b}\to \Lambda,p$ transition from factors, which challenges the naive power counting of SCET \cite{Han:2022srw,Feldmann:2011xf}. It implies that the QCD dynamics is indeed different between baryonic and mesonic systems, and that the power suppression of $1/m_{b}$ is not strictly respected. As stressed in Ref.\cite{Han:2022srw}, the leading power diagrams with two hard collinear gluons exchanging are largely suppressed by $\mathcal{O}(\alpha^{2}_{s})$ due to the fact that a baryon contains three valance quarks. Therefore, it is necessary to develop an insightful and comprehensive treatment on $b-$baryon non-leptonic decays in order to further improve our understandings. \textcolor{red}{We present the first study of nonleptonic two-body decays of bottom baryons under the \textcolor{red}{the re-scattering approach}, thereby predicting observables such as branching ratios and CP violation in bottom baryon decays.}}

\deleted{The QCD dynamics in baryon non-leptonic decays are very complicated and laborious to calculate, and various methods are developed in order to finish this task. Compared with the case of meson decays, the non-factorizable contribution and charm penguin are involved more frequently in the baryon decays. Although they are expected to be suppressed \cite{Leibovich:2003tw,Mantry:2003uz,Cheng:1996cs,Mohanta:2000nk,
Cheng:1992gv}, their effects are proven to be critical in the $B$ meson $CPV$ and can not be simply neglected \cite{Lu:2000em,Keum:2000wi,Beneke:2003zv}. Hence, it is definitely required to well estimate these contributions for reliable predictions about $\Lambda^{0}_{b}$ decays. A theoretical approach of QCD factorization based on collinear factorization \cite{Zhu:2018jet} is developed to study the $b-$baryon hadronic decays, while its approximation under the diquark hypothesis encounters serious challenges and need to be further tested in charmless $\Lambda^{0}_{b}$ decays. There are also some research on relations between decay rates and $CP$ asymmetries with the $SU(3)$ flavor symmetry \cite{Roy:2019cky,He:2015fwa,He:2015fsa}. Generalised factorization approach  is employed to make some predictions for $\Lambda^{0}_{b}\to p\pi^{-}/K^{-}/\rho^{-}/K^{*-}$ and $\Lambda\phi/\eta$ decays, where the NLO improved effective Hamiltonian and color parameter $N_{c}$ is used to incorporate the possible non-factorizable contributions \cite{Geng:2016gul,Hsiao:2015nvu,
Hsiao:2017tif}. In \cite{Cheng:1996cs}, many Cabibbo-allowed two-body hadronic weak decays of bottom baryons are analyzed within the naive factorization approach by using form factors under nonrelativistic quark model. PQCD is used to calculate $\Lambda^{0}_{b}\to p\pi^{-}/K^{-}$ in Ref.\cite{Lu:2009cm} and improved in Ref.\cite{Yu:2024cjd} by adding the high twist contributions. It is well known that a complete prediction based on the factorization schemes requires some non-perturbative inputs, such as the light cone distribution amplitudes (LCDAs) of hadrons whose determination with the first principle calculation of QCD is full of challenges. Particularly, the multibody $\Lambda^{0}_{b}$ decays are always calculated in quasi two body picture and many intermediate resonances like $N^{*}(1520),\Delta,\Lambda(1520)$... need to be included. Precise predictions are hard to made because of the limited knowledge about related LCDAs. However, it is these multibody decays that have remarkable potential in probing $b$-baryon $CPV$s.}

\deleted{The important role of final state interactions (FSIs) are easily accepted in hadronic charm decays due to the relatively small energy release and the fact that there exist many resonances nearby this energy scale \cite{Elaaoud:1999pj,Li:2020qrh,Jiang:2018oak,Li:2018epz,Cheng:2002wu,Han:2021gkl,Han:2021azw,Yu:2017zst, Ablikim:2002ep,Li:2002pj,Jia:2024pyb}. In bottom charmless decays the conventional viewpoint is that FSIs are expected to play a \replaced{There's a problem here }{mirror} role due to the large energy release. Nevertheless, in the non-leptonic $B$ meson decays \cite{Cheng:2004ru} it is indicated that soft FSIs might have some essential effects. Meanwhile, the FSIs in baryon system also worth to be investigated. We would like to talk more about the final states rescattering method as follows. }


In this work, we apply the final state re-scattering approach with hadronic loops for the first time to $\Lambda^{0}_{b}$ non-leptonic decays. We take the methodology in Ref.~\cite{Jia:2024pyb} to calculate triangle diagrams. Differently, we utilize form factors that are free from unphysical poles in the loop integrals to ensure the reliability of $CP$ asymmetry predictions. We calculate five two-body non-leptonic decays of \( b \)-baryons, with the aim of facilitating experimental searches for additional $CP$-violating baryon decay processes.


%

\deleted{In this work, we employ the FSIs with loop mechanism to study five representative charmless decays of $\Lambda^{0}_{b}$. \replaced{We look forward to future experiments to verify our computational results}{We show that this approach works well in b-baryon decays.} Firstly we fix two parameters with the help of experimental data of $\Lambda^{0}_{b}\to p\pi^{-},pK^{-}$, and make predictions for the other observables. Our method to study b-baryon decays in this work can be applied to calculation of more decays. In section II, the theoretical framework is introduced, including the introduction of effective Hamiltonian, topological diagrams of baryon weak decays, short distance contributions described by the naive factorization hypothesis, long distance dynamics estimated using FSIs. We investigate the helicity amplitudes in section , branching ratios, the direct CP asymmetries and asymmetry parameters. The complex formulations are collected in the appendixes. The effective Lagrangian and some cumbersome formula such as loop amplitudes are listed in the Appendix~\ref{app.A},\ref{app.B}, and \ref{app.C}.}

The paper is organized as follows. In Section \ref{framework}, we introduce the theoretical framework, which includes the effective Hamiltonian, topological diagrams of baryon weak decays, and both short- and long-distance amplitudes. The helicity amplitudes are given in Section \ref{amplitudes} . In Section \ref{ppi-pk}, we discuss the branching ratios, direct $CP$ asymmetries, and asymmetry parameters for  $\Lambda^{0}_{b}\to p\pi^{-}$ and $pK^{-}$. The results and discussion about $\Lambda^{0}_{b}\to p\rho^{-}$, $pK^{*-}$ and $\Lambda\phi$ decays are presented in Section \ref{prhoklambdaphi}, and the Section \ref{Summary} is a summary. The effective strong Lagrangian, strong coupling vertices,  the analytical expressions of the amplitudes,and the full amplitudes of the five decay channels are collected in Appendices \ref{app.A}, \ref{app.Bprime}, \ref{app.B}, and \ref{app.C}, respectively.
\section{Theoretical framework}\label{framework}

In this section, we introduce the topological diagrams, the naive factorization estimation of short-distance amplitudes, and the re-scattering picture for long-distance contributions.

\subsection{\deleted{Effective Hamiltonian and}Topological diagrams}

\deleted{There are three separated energy scales in the heavy b-baryon decays, i.e., $m_{W}\gg m_{b}\gg\Lambda_{QCD}$, and an efficient and conventional idea is effective field theory by integrating out some heavy degrees of freedom, and then performing local operator product expansion to get a series of local composite operators. After this, all high energy information above scale $m_{b}$ are integrated into Wilson coefficients which \textcolor{red}{are}\deleted{is} independent from specific external states, and therefore can be determined perturbatively order by order through matching at quark level. }

The charmless two-body non-leptonic  weak decays of bottom baryons are governed by the $b\to u$ transition at tree level and the $b\to q$ transition (with $q = d,s$) at loop level. These quark-level decays are typically calculated using effective field theory, with an effective Hamiltonian given by \cite{Buchalla:1995vs}
\begin{equation}
	\begin{aligned}
    \mathcal{H}_{eff}= & \frac{G_F}{\sqrt{2}}\left\{V_{u b} V_{u q}^*\left[C_1(\mu) \mathcal{O}_1^u(\mu)+C_2(\mu) \mathcal{O}_2^u(\mu)\right]-V_{t b} V_{t q}^*\left[\sum_{i=3}^{10} C_i(\mu) \mathcal{O}_i(\mu)\right]\right\}+ \text { H.c. }
	\end{aligned}
\end{equation}
where $C_{i}~(i=1,...,10)$ are Wilson coefficients evaluated at renormalization scale $\mu=m_b$ and $\mathcal{O}_{i}~(i=1,...,10)$ are four quark   operators. 

\begin{figure}[htbp]
\centering
\includegraphics[height=10cm,width=11.5cm]{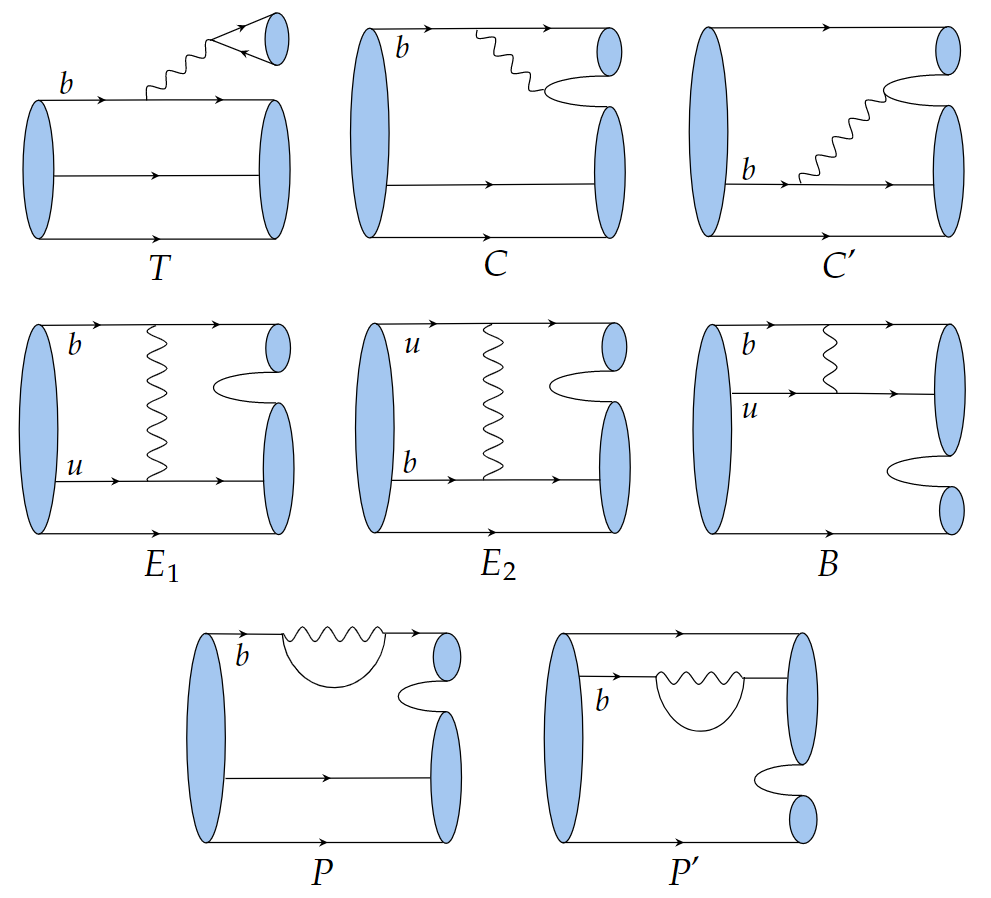}
\caption{The topological diagrams for two body hadronic decays of $\Lambda^{0}_{b}$ baryon, where the first two rows are tree diagrams and the third row are penguin type diagrams.}
\label{topological diagrams}
\end{figure}

The theoretical realization of non-leptonic hadron decays involves calculating \deleted{is to calculate the} matrix element of these local effective operators with definite external states, and the topological diagrams are viewed as an intuitive representation of these elements that involve all possible strong dynamics including both perturbative and non-perturbative parts \cite{Cheng:2004ru}. As depicted in Fig.\ref{topological diagrams}, we present \deleted{introduce} all possible topological diagrams for $\Lambda^{0}_{b}$ decays, which are sorted according to the typologies of weak vertex. Specifically, they are
\begin{itemize}
    \item $T$: color-allowed diagram with external $W$-emission.
    \item $C$ and $C^{\prime}$: color-suppressed internal $W$-emission diagrams, where the difference between them is that the quark generated from bottom quark weak decay flows into the final-state meson ($C$) or baryon ($C^\prime$).
    \item $E_{1}$, $E_{2}$ and $B$: three distinct types of $W$-exchange diagrams, distinguished by the flow of quarks produced from the weak vertex.
    \item $P$ and $P^{\prime}$: two types of diagrams with penguin operators.
\end{itemize}

Although the topological diagrams are classified by the structure of weak vertices, they also involve all the strong interaction dynamics of both perturbative and nonperturbative effects. The $T$ amplitude can be easily proved factorization in the soft collinear effective theory \cite{Leibovich:2003tw}, thus it can be effectively estimated using the naive factorization approach, under which it can be expressed as the product of baryon weak transition form factors and meson decay constants. Under the naive factorization, $W$ internal emission $C$ and exchanging $E$ are expected to be largely suppressed in $B$ meson decays by the small Wilson coefficient $a_{2}$, and thus are insufficient to explain the experimental data for $C,E$ dominated processes \cite{Cheng:2004ru}. 
Meanwhile,  extraction of amplitude ratios $C/T$ and $E/T$ in $B$ meson decays from the experimental data indicated that non-factorizable long-distance contributions \deleted{of them} play a significant role \cite{Cheng:2004ru}. The calculation of $B$ -decays in QCDF implies $a_{2}\approx 0.2$ by taking the hard spectator contribution into account, and hence gains a large enhancement compared to naive estimation \cite{Beneke:2001ev}. The power counting rules derived from the soft-collinear effective theory give the ratios among the different topological diagrams as $\frac{|C|}{|T|}\sim \frac{|C^{\prime|}}{|C|}\sim \frac{|E_{1}|}{|C|}\sim \frac{|E_{2}|}{|C|}\sim \frac{|B|}{|C|}\sim \mathcal{O}(\frac{\Lambda_{QCD}}{m_{b}})$, which implies the contributions from $C$ and $E$ diagrams can be suppressed \deleted{ignored} in $b$-decays \cite{Leibovich:2003tw,Mantry:2003uz}. With these counting rules one can safely obtain an estimation for the branching ratios without worrying about the non-factorizable contributions. However, when one studies the CPV, their effects become very critical and should be calculated reliably. In this work we reach this goal by considering long-distance FSIs effects.

\subsection{Short distance contributions under the factorization hypothesis}

In this subsection, we will give a concise introduction to the naive factorization approach for estimating the short distance contributions of $T,C,P$ diagrams. The decay amplitudes of $\mathcal{B}_{b}\to \mathcal{B}M$ is generally given as
    \begin{equation}\label{eq.1}
	\begin{aligned}
\bra{\mathcal{B}M}\mathcal{H}_{eff}\ket{\mathcal{B}_{b}}=\frac{G_{F}}{\sqrt{2}}V_{CKM}\sum_{i}C_{i}\bra{\mathcal{B}M}\mathcal{O}_{i}\ket{\mathcal{B}_{b}} .
	\end{aligned}
\end{equation}
With the naive factorization, the associated amplitudes can be expressed as the product of two parts: the decay constant of meson $M$ and heavy to light baryonic form factors.\deleted{ We take $\Lambda^{0}_{b}\to p\pi^{-}$ and $p\rho^{-}$ as examples. The general form of their amplitudes are classified into $\mathcal{B}_{b}(p_{i})\to \mathcal{B}(p_{f})P$ and $\mathcal{B}_{b}(p_{i})\to \mathcal{B}(p_{f})V$ respectively, which can be parameterized as \cite{Lee:1957qs,Pakvasa:1990if}.}  
Generally, the amplitudes for $\mathcal{B}_{b}(p_{i})\to \mathcal{B}(p_{f})P$ and $\mathcal{B}_{b}(p_{i})\to \mathcal{B}(p_{f})V$ are parameterized as\cite{Pakvasa:1990if}:
\begin{equation}
	\begin{aligned}
       \mathcal{A}(\mathcal{B}_{b}(p_{i})\to \mathcal{B}(p_{f})P)=i\bar{u}(p_{f})\left[A+B\gamma_{5}\right]u(p_{i}),
	\end{aligned}\label{Pesudo}
\end{equation}
\begin{equation}
	\begin{aligned}
       \mathcal{A}(\mathcal{B}_{b}(p_{i})\to \mathcal{B}(p_{f})V)=\bar{u}(p_{f})\left[A_{1}\gamma_{\mu}\gamma_{5}+A_{2}\frac{p_{f,~\mu}}{M_{i}}\gamma_{5}+B_{1}\gamma_{\mu}+B_{2}\frac{p_{f,~\mu}}{M_{i}}\right]\epsilon^{*\mu}u(p_{i}),
	\end{aligned}\label{Vector}
\end{equation}
where $P$ and $V$ are pseudoscalar and vector meson, respectively, $u(p_{i}),{u}(p_{f})$ are the Dirac spinors of initial $\mathcal{B}_{b}(p_{i})$ and final $\mathcal{B}(p_{f})$, and $\epsilon^{\mu}$ is the polarization vector of final $V$. The parameters $A$, $B$, $A_{1},A_{2}$ and $B_{1},B_{2}$ are derived within naive factorization as

\begin{equation}
	\begin{aligned}
      A&=\lambda_A f_{P}(M_{i}-M_{f})f_{1}(q^{2}),\\
      B&=\lambda_B f_{P}(M_{i}+M_{f})g_{1}(q^{2}),\\
     A_{1}&=-\lambda m_{V}f_{V}\left[g_{1}(q^{2})+\frac{M_{i}-M_{f}}{M_{i}}g_{2}(q^{2})\right],\\
    B_{1}&=\lambda m_{V}f_{V}\left[f_{1}(q^{2})-\frac{M_{i}+M_{f}}{M_{i}}f_{2}(q^{2})\right],\\
         A_{2}&=-2\lambda m_{V}f_{V}f_{2}(q^{2}),\\
    B_{2}&=2\lambda m_{V}f_{V}g_{2}(q^{2}),
	\end{aligned}
\end{equation}
where $M_{i},M_{f}$ are masses of initial and final baryons, respectively. $f_P$ and $f_V$ are the decay constants of pseudoscalar and vector mesons, respectively.  $f_1,f_2,g_1$ and $g_2$ denote the heavy-to-light transition form factors in $\Lambda_b^0$ decays. In our work, we will use the results from Ref. \cite{Zhu:2018jet}, where the form factors for $\Lambda^{0}_{b} \to p, n, \Lambda^{+}_{c}, \Lambda$  are derived under a uniform model.  $\lambda_A$,$\lambda_B$ and $\lambda$ functions are process dependent and incorporate both CKM factors and Wilson coefficients, which are listed in Appendix C of Ref.\cite{Zhu:2018jet}.

\deleted{The results for the other decays are also derived and collected in Ref.\cite{Zhu:2018jet}. The short-distance amplitudes are also essential in understanding the long-distance contributions, as can be seen in the following discussion.}

\subsection{Long-distance contributions with the re-scattering mechanism}

The nonfactorization contributions of color-suppressed $C$ and $W$-exchange $E$ topology graphs, which account for the relative strong phases, are important for predicting CP asymmetries.
Final-state re-scatterings provide a natural physical picture of the long-distance contributions in heavy hadron decays. Wolfenstein and Suzuki proposed a formalism for final-state interactions at the hadron level, based on $CPT$ invariance and unitarity \cite{Wolfenstein:1990ks,Suzuki:1999uc}. A comprehensive study was performed on $B$-meson two-body decays to examine the $B$ decay rates and their impacts on direct $CP$ asymmetries by incorporating FSI effects \cite{Cheng:2004ru}. Employing the time evolution picture of scatterings, the short-distance interactions occur rapidly and violently at the beginning of weak decays, while the long-distance ones take place at a much later time. The full amplitude is expressed as \cite{Cheng:2020ipp}
\begin{equation}\label{re-scattering}
	\begin{aligned}
\mathcal{A}(\Lambda^{0}_{b}\to f)=\sum_{i}\bra{f}U(+\infty,\tau)\ket{i}\bra{i}\mathcal{H}_{eff}\ket{\Lambda^{0}_{b}} ,
	\end{aligned}
\end{equation}
where $\tau$ is a very short time interval characterizing the weak decay scale. Within the naive factorization framework, the matrix element 
$\bra{i}\mathcal{H}_{eff}\ket{\Lambda^{0}_{b}}$ does not develop strong interaction phases. The re-scattering part $\bra{f}U(+\infty,\tau)\ket{i}$ introduces a complex amplitude with non-zero phase, just as $\pi\pi\to KK$ inelastic scattering in the $B$ meson three-body decays \cite{Cheng:2020ipp,Garrote:2022uub,AlvarengaNogueira:2015wpj,Bediaga:2013ela}. It has been manifested that the final-state re-scatterings are very important for the CPV of three-body $B$ meson decays.

\begin{figure}[htbp]
\centering
\includegraphics[height=8cm,width=16cm]{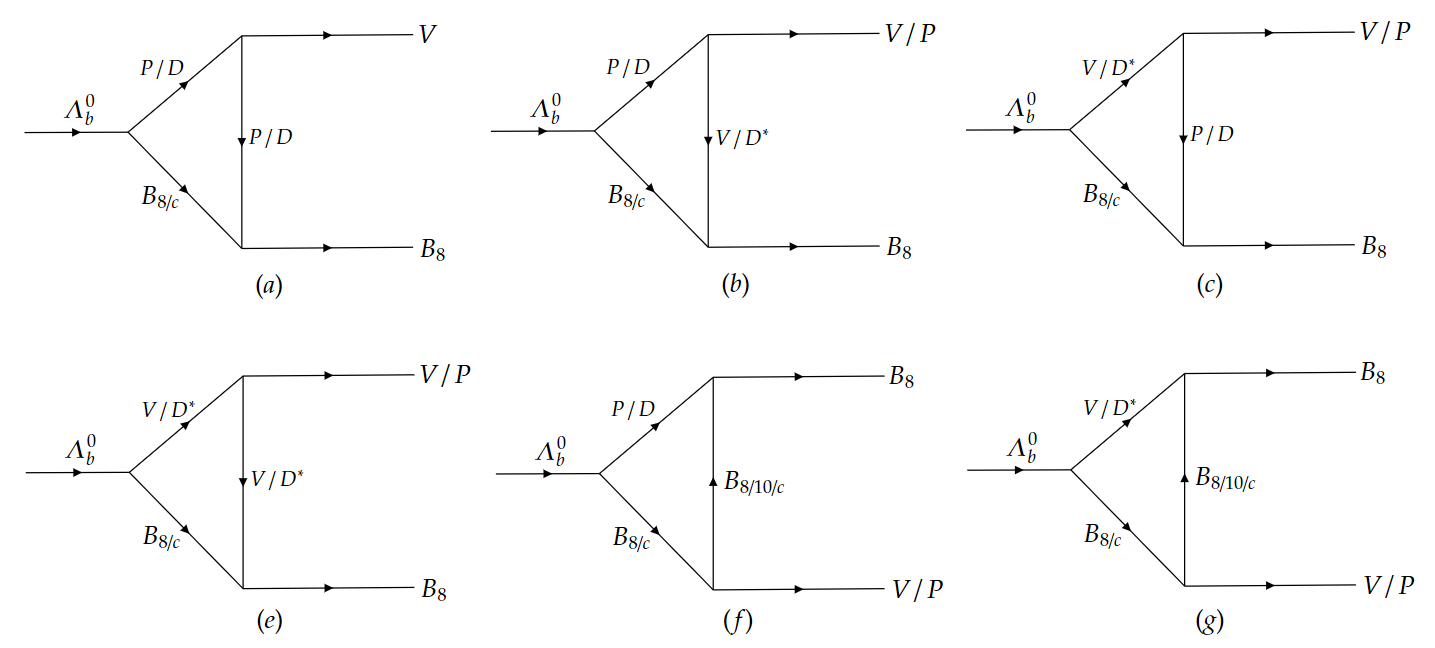}
\caption{The long-distance re-scattering contributions of $\Lambda^{0}_{b}$ decays at hadron level under single-particle exchange approximation, where the $B_{b},B_{c},B_{8},B_{10}$ denote bottom, charm, octet and decuplet baryons, and $P,V,D,D^{*}$ are pseudo-scalar octet and vector, $D$ and $D^{*}$ mesons, respectively.}
\label{triangle diagrams}
\end{figure}

Estimation of these non-factorization effects is challenging due to their non-perturbative nature. Nevertheless, they can be estimated using the single particle exchange approximation at the hadron level. Specifically, the strong scattering matrix element $\bra{f}U(+\infty,\tau)\ket{i}$ is treated as the re-scatterings of two intermediate hadrons following $\Lambda^{0}_{b}$ weak decays. Under this mechanism, the long-distance contributions to $\Lambda^{0}_{b}$ two-body hadronic decays are described by the triangle diagrams as shown in Figs.~\ref{triangle diagrams} and \ref{anni diagrams}. In order to calculate these triangle diagrams, one needs to combine the derivation of the weak vertex, treated as short-distance amplitudes under naive factorization, with hadronic re-scatterings governed by the effective Lagrangian collected in Appendix~\ref{app.A}. The associated Feynman rules for strong vertices are obtained by inserting effective operators for specific initial and final hadron states and listed in Appendix~\ref{app.Bprime}.

Then, one can get the analytical amplitudes of triangle diagrams by a loop integral with these weak and strong vertices as well as hadron propagators. Next, we use an example to illustrate our derivation and the conventions of symbols. Considering the decay of $\Lambda^{0}_{b}(p_{i},{\lambda}_{i})\to p(p_{3},\lambda_{3})\rho^{-}(p_{4},\lambda_{4})$, with intermediate particles $B(p_{2},\lambda_{2})P(p_{1})$ re-scattering via exchanging $V(k,\lambda_{k})$, as depicted in diagram $(b)$ of Fig.\ref{triangle diagrams}. Its final analytical amplitude is expressed as an integral over the inner momentum $k$
\begin{equation}
 \begin{aligned}
\mathcal{M}[P_8,B_8;V]
     &=\int\frac{d^4k}{(2\pi)^4}\frac{4g_{P_8VV}}{f_{P_8}}\bar{u}(p_4,\lambda_4)(f_{1VB_8B_8}\gamma_\delta-\frac{if_{2VB_8B_8}}{m_2+m_4}\sigma_{\rho\delta}k^\rho)(\not\!p_2+m_2)(A+B\gamma_5)u(p_i,\lambda_i)\\
     &\times(-g^{\delta\nu}+\frac{k^\delta k^\nu}{m_k^2})
     \varepsilon^{\mu \nu \alpha \beta}\varepsilon_{\beta}^*(p_3,\lambda_3)k_{\mu}p_{3\alpha}\cdot\frac{\mathcal{F}}{(p_1^2-m_1^2+i\varepsilon)(p_2^2-m_2^2+i\varepsilon)(k^2-m_k^2+i\varepsilon)}.
         \end{aligned}
\end{equation}
We use the label $\mathcal{M}[P_8, B_8; V]$ to represent the amplitude of the scattering between \deleted{of} an octet pseudoscalar meson ($P_8$) and an octet baryon ($B_8$) by exchanging a vector meson $V$. The amplitude expressions for all triangle diagrams are listed in Appendix~\ref{app.B}. \deleted{In Appendix.\ref{app.B}, expressions of all triangle amplitudes used in this work are reorganized exclusively.} The strong couplings for 3-hadron vertex are derived at on shell hadrons, thus form factor $\mathcal{F}$ is introduced to take care of the off-shell and sub-structure effects of exchanged particles, and meanwhile, regularize the potential UV divergence in the loop integral. Here, we adopt the following model \cite{Yue:2024paz}
\begin{equation}\label{FF}
\mathcal{F}(\Lambda, m_k) = \frac{\Lambda^4}{(k^2 - m_k^2)^2 + \Lambda^4},
\end{equation}
where the $\Lambda$ is a model parameter.  In the recent work on charm baryon decays, many $\Lambda^{+}_{c} \to BV$ decaying channels have been explained and predicted using one model parameter $\eta$, assuming $SU(3)$ flavor symmetry \cite{Jia:2024pyb}. This is because all re-scattering and final state particles are light and located nearly on the same energy scale.   For $b$-baryon decays, it is however not a sensible prescription, as the charmed hadronic rescatterings associated with charm loop effects have to be taken into account for a reasonable treatment of $CP$ asymmetries,   as investigated in the $B^{\pm} \to \pi^{+}\pi^{-}\pi^{\pm}$ high mass region \cite{Bediaga:2020ztp} . It makes sense that the regularization parameters $\Lambda$ are not universal in scatterings of light particles like $p\pi^{-} \to p\pi^{-}$ and the scatterings of charmed heavy particles like $\Lambda^{+}_{c}D^{-} \to p\pi^{-}$. Hence, we employ two different parameters, namely $\Lambda_{\textbf{charmless}}$ and $\Lambda_{\textbf{charm}}$, to characterize these two distinct scattering modes. Both parameters will be determined by using the experimental data of branching ratios and $CP$ asymmetries of $\Lambda^{0}_{b} \to p\pi^{-}$ and $pK^{-}$. We make some comments as follows.
\begin{itemize}
    \item In principle, the full \deleted{whole} amplitudes $\mathcal{A}(\Lambda^{0}_{b}\to f)$ in Eq.\eqref{re-scattering} should be treated by reorganizing contributions from all symmetry allowed intermediate states as required by unitarity. However, it is highly challenging to incorporate multi-body hadronic scatterings since they are very complicated and in general not under theoretical control. Therefore, we first assume that the dominant contributions arise from \deleted{it is prominent in} the $2\to 2$ processes, and thus view this treatment as a working tool as in \cite{Cheng:2004ru}. We work out the consequences of this tool to see if it is empirically working.
    \item The regulator form factor in Eq.\eqref{FF} contains no \deleted{is free from} potential poles, and thus the imaginary part of amplitudes in our calculation is completely induced by physical states in loops according to Cutkosky's rule \cite{Peskin:1995ev}. Hence, there are \deleted{hence} no unphysical strong phases \deleted{are} introduced by \deleted{involved under}this regulator.
    \item In our work, the bubble contribution, as depicted in Fig.\ref{anni diagrams}, will be ignored since it is expected to be suppressed relative to that arising from triangle diagrams due to the lack of resonances near the threshold of the $b$-baryon mass, although it plays an important role in charm decays \cite{Cheng:2004ru}.
    
    \begin{figure}[htbp]
    \centering
   \includegraphics[height=4cm,width=7cm]{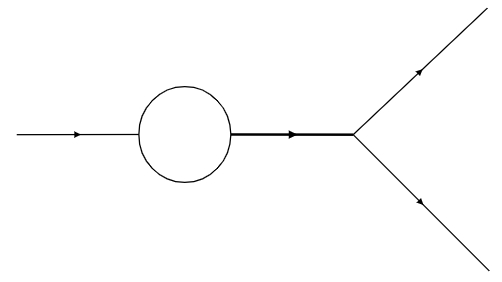}
   \caption{The long distance contribution through a resonance propagation. }
   \label{anni diagrams}
   \end{figure}
    \item The contributions with charmonium intermediate states, for example, $\Lambda J/\psi\to pK^{-}$, are largely suppressed owing to the small Wilson coefficients at the \deleted{in} weak vertex and the negligible strong couplings for \deleted{of} $J/\psi\to p\bar{p}$ \cite{Fu:2023ose,He:2022jjc}. Hence, we ignore them in this work.
\end{itemize}
\section{Helicity Amplitudes }\label{amplitudes}

In this section, we first present the input parameters for numerical calculations. The helicity amplitudes of five $\Lambda^{0}_b$ decay channels are then evaluated by considering both short- and long-distance contributions. Finally,   phenomenological discussions are presented.

\subsection{Input parameters}
The baryon masses we employed are
$m_{\Lambda^{0}_{b}}=5.619$ GeV, $m_{\Lambda^{+}_{c}}=2.286$ GeV, $m_{p}=0.938$ GeV, and meson masses $m_{D}=1.869$ GeV, $m_{\pi}=0.140$ GeV, $m_{K}=0.490$ GeV, $m_{\rho}=0.770$ GeV, $m_{K^{*}}=0.892$ GeV. The quark masses are current masses. Here, we take the values as $m_{u}=2.16$ MeV, $m_{d}=4.70$ MeV, $m_{s}=93.5$ MeV, $m_{c}=1.27$ GeV, $m_{b}=4.18$ GeV \cite{ParticleDataGroup:2022pth}.

The CKM quark-mixing matrix elements are adopted under Wolfenstein parameterization with leading expansion
$V_{ud}=1-\frac{\lambda^{2}_{W}}{2}, V_{us}=\lambda_{W}, V_{ub}=A\lambda^{3}_{W}(\rho-i\eta)$, $V_{cd}=-\lambda_{W},V_{cs}=1-\frac{\lambda^{2}_{W}}{2}, V_{cb}=A\lambda^{2}_{W}$ and $V_{td}=A\lambda^{3}_{W}(1-\rho-i\eta), V_{ts}=-A\lambda^{2}_{W},V_{tb}=1$, where the Wolfenstein parameters are taken as $A=0.823, \rho=0.141,\eta=0.349$ and $\lambda_{W}=0.225$. Here, we sign $\lambda_{W}$ to distinguish it from $\lambda$ functions we used before \cite{ParticleDataGroup:2022pth}.

The heavy-to-light form factors for $\Lambda^{0}_{b} \to p, n, \Lambda, \Lambda^{+}_{c}$ are used as inputs in our calculation, which have been extensively investigated. We use the data of form factors in Ref.\cite{Zhu:2018jet}. The decay constants of pseudoscalar and vector mesons are listed in  Table~\ref{tab:decay_constant}, where the definition of $f_{\eta^{(\prime)}_{u,d,s}}$ is the same as that in Ref.\cite{Zhu:2018jet}.
\begin{table}[htbp]
\renewcommand{\arraystretch}{1.5}
\addtolength{\arraycolsep}{3pt}
	\centering
\caption{The decay constants of pseudoscalar and vector mesons used in this work \cite{Zhu:2018jet,Cheng:2004ru}.} \label{tab:decay_constant}
	\begin{tabular}{c|c|c|c|c|c|c|c|c|c} 
		\hline
		\hline
Decay constant &  $f_\pi$ & $f_K$  &  $f_\rho$  & $f_\omega$  & $f_\phi$ &  $f_{K^*}$  & $f_{D^{*}}$ & $f_{D}$ & $f_{D_{s}}$ \\
\hline
Value [MeV] &  $ 130.3 $ & $ 155.7 $ &  $ 216$ & $  187$ & $ 215 $ & $210 $ & $230$ & $212$ & $250$\\
\hline
\hline
Decay constant &  $f_{D^{*}_{s}}$ & $f_{J/\psi}$  &  $f_{\eta_{c}}$  & $f_{\eta_{u}}$  & $f_{\eta_{d}}$ &  $f_{\eta_{s}}$  & $f_{\eta^{\prime}_{u}}$ & $f_{\eta^{\prime}_{d}}$ & $f_{\eta_{s}^{\prime}}$ \\
\hline
Value [MeV] &  $ 271 $ & $ 418 $ &  $ 387$ & $  54$ & $ 54 $ & $-111 $ & $44$ & $44$ & $136$\\ 
		\hline
		\hline 
	\end{tabular} 
\end{table}
\\

In addition, the strong coupling constants serve as \deleted{are} crucial non-perturbative parameters. For \deleted{the strong} couplings between octet baryons and light mesons, we adopt \deleted{use their} values obtained via light-cone sum rules (LCSRs) \deleted{with the method of LCSRs }under the $SU(3)$ flavor symmetry \cite{Aliev:2006xr,Aliev:2009ei}. The couplings involving \deleted{among} decuplet baryons, octet baryons and light mesons are extracted \deleted{obtained with the } from experiment data:\deleted{ that are} $g^{2}_{\Delta N\pi}/(4\pi)=0.36$ and $g^{2}_{\Delta N\rho}/(4\pi)=20.45$ \cite{Janssen:1996kx}, with remaining couplings \deleted{of this type of couplings are } determined through \deleted{with the} $SU(3)$ flavor symmetry. Three kinds of meson couplings are determined by using the values of three representative couplings $g_{\rho\pi\pi}$, $g_{\rho\rho\rho}$, $g_{\omega\rho\pi}$ under the $SU(3)$ flavor symmetry. The on-shell coupling constant $g_{\rho\pi\pi} = 6.05$ is determined by the decay rate of $\rho\to \pi\pi$. The hidden local symmetry theory \cite{Meissner:1987ge} relates the $\rho\rho\rho$ coupling constant $g_{\rho\rho\rho}$ to the $\rho$ meson mass, given by $g_{\rho\rho\rho} = \frac{m_\rho}{2 F_\pi}$ with $F_\pi = \frac{f_\pi}{\sqrt{2}}$.  Additionally, the coupling constant $g_{\omega \rho \pi}$ can be expressed as $g_{\omega \rho \pi} = \frac{3}{16 \pi^2} g_{\rho\rho\rho}^2$, describing the interaction between the $\omega$ meson, $\rho$ meson, and a pion. The strong couplings of two charm mesons and a light meson  are given as \cite{Cheng:2004ru}
$$g_{D^*D^*P_8}=\frac{g_{D^*DP_8}}{\sqrt{m_D m_{D^*}}},\quad g_{DDV}=g_{D^*D^*V}=\frac{\beta g_V}{\sqrt{2}},\quad f_{D^*DV}=\frac{f_{D^*D^*V}}{m_{D^*}}=\frac{\lambda g_V}{\sqrt{2}},$$
where $g_V=\frac{m_{\rho}}{f_\pi},\quad \beta=0.9,\quad 
\lambda=0.56\mbox{ GeV}^{-1}$ and $g_{D^*DP_8}=17.9$ extracted from the experimental data of $D^{*}$ width. Finally, the strong couplings described charm baryons, charm mesons and light octet baryons can be found in Ref.\cite{Yu:2017zst}. The effective Lagrangians to describe strong scatterings at hadron level are collected in Appendix.\ref{app.A}.

\subsection{Numerical results}

As introduced in eq.(\ref{FF}),  the  form factor $\mathcal{F}(m_{k},\Lambda)$ is introduced to characterize   the   off-shell level of intermediate resonances and regulate possible divergences in loop integrals. We determine these model parameters as $\Lambda_{\textbf{charm}}=1.0$ and $\Lambda_{\textbf{charmless}}=0.5$ from the experimental data of branching ratios  and CP asymmetry of $\Lambda^{0}_{b}\to p\pi^{-}$ and $pK^{-}$ decays \cite{ParticleDataGroup:2022pth}.

Using \deleted{With} above model parameters, we perform our calculations in the helicity basis where \deleted{under which} the asymmetry parameters are more readily defined \deleted{easier to be defined} \cite{Chen:2019hqi,Hong:2022prk}. We specifically present helicity amplitudes for five decay channels,  which can be tested by experiments after the partial wave analysis is implemented in the future. Table~\ref{Helicity Lb to pK} shows $\Lambda^{0}_{b} \to pK^{-}$ results as an illustrative example, where $\mathcal{S}$, $\mathcal{NC}$, and $\mathcal{C}$ represent short-distance amplitudes, re-scattering amplitudes with charmless loop and charm triangle loop, respectively.  We also show the contributions with different CKM factors separately for comparison. The short-distance and charmless loop amplitudes of $\Lambda^{0}_{b} \to pK^{-}$ contain both tree $(V_{ub}V^{*}_{us})$ and penguin $(V_{tb}V^{*}_{ts})$ components. The charm triangle loops, which are associated with $V_{cb}V^{*}_{cs}$, incorporate the re-scatterings of $\Lambda^{0}_{b}\to \Lambda^{+}_{c}D^{(*)}_{s} \to pK^{-}$, and in fact recognize the long-distance contributions of charm penguin diagrams as $B\to D_{s}D\to K\pi$ for $B$ meson system \cite{Ciuchini:1997rj,Colangelo:1989gi}. We want to remind that the numerical results in Tables \ref{Helicity Lb to pK} to \ref{Helicity Lb to prho}   collecting helicity amplitudes of     five $\Lambda_b^0$ decay modes do not include the CKM matrix elements.

\begin{sidewaystable}
\caption{Helicity amplitudes of $\Lambda_b^0 \rightarrow pK^-(10^{-7})$ with different $CKM$ factors}
\centering
\begin{tabular}{ccccccc}
\toprule
\toprule
 decay modes & $H_{-\frac{1}{2}} ~(V_{ub}V_{us}^*) $ &$H_{-\frac{1}{2}} ~(V_{cb}V_{cs}^*)$ & $H_{-\frac{1}{2}} ~(V_{tb}V_{ts}^*)$ & $H_{\frac{1}{2}} ~(V_{ub}V_{us}^*) $ &$H_{\frac{1}{2}} ~(V_{cb}V_{cs}^*)$ & $H_{\frac{1}{2}} ~(V_{tb}V_{ts}^*)$ \\
\midrule
$\mathcal{S}(\Lambda_b^0 \rightarrow pK^-)$& 104.66 $i$ & --- & 3.06 $i$  & 21.07 $i$ & --- & 4.16 $i$  \\
$\mathcal{NC}(\Lambda_b^0 \rightarrow pK^-)$& -11.15 - 7.66 $i$ &--- & -0.42 - 0.29 $i$& -8.29 - 4.76 $i$&---&-0.59 - 0.38 $i$\\
$\mathcal{C}(\Lambda_b^0 \rightarrow pK^-)$& --- &-2.80 + 0.73 $i$ & 0.08 - 0.14 $i$ & ---&5.30 - 2.65 $i$&0.003 + 0.09 $i$\\
\bottomrule
\bottomrule
\end{tabular}\label{Helicity Lb to pK}

\vspace{1cm}

\caption{Helicity amplitudes of $\Lambda_b^0 \rightarrow p\pi^-(10^{-7})$}
\centering
\begin{tabular}{cccccccc}
\toprule
\toprule
 decay modes & $H_{-\frac{1}{2}} ~(V_{ub}V_{ud}^*) $ &$H_{-\frac{1}{2}} ~(V_{cb}V_{cd}^*)$ &$H_{-\frac{1}{2}} ~(V_{tb}V_{td}^*)$ & $H_{\frac{1}{2}} ~(V_{ub}V_{ud}^*) $ &$H_{\frac{1}{2}}~(V_{cb}V_{cd}^*)$ & $H_{\frac{1}{2}} ~(V_{tb}V_{td}^*)$ \\
\midrule
$\mathcal{S}(\Lambda_b^0 \rightarrow p\pi^-)$& 85.00 $i$ &--- & 2.56 $i$ &17.01 $i$ &---&3.73 $i$\\
$\mathcal{NC}(\Lambda_b^0 \rightarrow p\pi^-)$& -9.18 - 2.31 $i$&--- & 0.01 - 0.04 $i$ &3.11 - 4.56 $i$&---&-0.002 - 0.17 $i$\\
$\mathcal{C}(\Lambda_b^0 \rightarrow p\pi^-)$& ---& 0.53 - 1.78 $i$ & 0.15 - 0.17 $i$ & ---&8.74 - 6.90 $i$&0.11 - 0.04 $i$\\
\bottomrule
\bottomrule
\end{tabular}\label{Helicity Lb to ppi}

\vspace{1cm}

\caption{Helicity amplitudes of $\Lambda_b^0\rightarrow \Lambda \phi (10^{-7})$}
\centering
\begin{tabular}{cccccccc}
\toprule
\toprule
 decay modes & $H_{0,-\frac{1}{2}} ~(V_{ub}V_{us}^*) $ &$H_{0,-\frac{1}{2}} ~(V_{cb}V_{cs}^*)$&$H_{0,-\frac{1}{2}} ~(V_{tb}V_{ts}^*)$ & $H_{-1,-\frac{1}{2}} ~(V_{ub}V_{us}^*) $ &$H_{-1,-\frac{1}{2}} ~(V_{cb}V_{cs}^*)$&$H_{-1,-\frac{1}{2}} ~(V_{tb}V_{ts}^*)$ \\
\midrule
$\mathcal{S}(\Lambda_b^0 \rightarrow \Lambda \phi )$&---&---& 2.58  &---&--- 
 &0.75 \\
$\mathcal{NC}(\Lambda_b^0 \rightarrow \Lambda \phi )$&4.81 - 8.94 $i$&---& 0.13 -0.21 $i$&1.76 - 1.86 $i$ &---&0.21 - 0.25 $i$\\
$\mathcal{C}(\Lambda_b^0 \rightarrow \Lambda \phi )$&---& 5.41 + 2.12 $i$&0.10 -0.02 $i$&---&2.11 + 2.00 $ i$& 0.09 + 0.15 $i$\\
\midrule
& $H_{1,\frac{1}{2}} ~(V_{ub}V_{us}^*) $ &$H_{1,\frac{1}{2}} ~(V_{cb}V_{cs}^*)$&$H_{1,\frac{1}{2}} ~(V_{tb}V_{ts}^*)$ & $H_{0,\frac{1}{2}}~(V_{ub}V_{us}^*) $ &$H_{0,\frac{1}{2}} ~(V_{cb}V_{cs}^*)$&$H_{0,\frac{1}{2}}~(V_{tb}V_{ts}^*)$ \\
\midrule
$\mathcal{S}(\Lambda_b^0 \rightarrow \Lambda \phi )$& --- & --- & - 0.03&---&---& - 0.01\\
$\mathcal{NC}(\Lambda_b^0 \rightarrow \Lambda \phi )$&5.45 - 5.75 $i$&---&0.17 -0.17 $i$&1.17 - 7.99 $i$&---&-0.01 - 0.16 $i$\\
$\mathcal{C}(\Lambda_b^0 \rightarrow \Lambda \phi )$&---&-5.35 - 1.67 $i$&-0.10 + 0.03 $i$&---& 0.84 - 1.22 $i$& - 0.03+0.01 $i$\\
\bottomrule
\bottomrule
\end{tabular}\label{Lb to Lambdaphi}
\end{sidewaystable}

\begin{sidewaystable}[htbp]
\caption{Helicity amplitudes of $\Lambda_b^0\rightarrow pK^{*-} (10^{-7})$}
\centering
\begin{tabular}{cccccccc}
\toprule
\toprule
 decay modes & $H_{0,-\frac{1}{2}} ~(V_{ub}V_{us}^*) $ &$H_{0,-\frac{1}{2}} ~(V_{cb}V_{cs}^*)$&$H_{0,-\frac{1}{2}} ~(V_{tb}V_{ts}^*)$ & $H_{-1,-\frac{1}{2}}~(V_{ub}V_{us}^*) $ &$H_{-1,-\frac{1}{2}} ~(V_{cb}V_{cs}^*)$&$H_{-1,-\frac{1}{2}} ~(V_{tb}V_{ts}^*)$ \\
\midrule
$\mathcal{S}(\Lambda_b^0 \rightarrow pK^{*-})$& 138.21 & --- & 3.06  & 31.58 & --- & 0.70  \\
$\mathcal{NC}(\Lambda_b^0 \rightarrow pK^{*-})$& -13.54 + 25.90 $i$ &--- & -0.36 + 0.58 $i$  & -2.78 + 4.77 $i$ &---& -0.26 + 0.30 $i$\\
$\mathcal{C}(\Lambda_b^0 \rightarrow pK^{*-})$& --- & - 9.23 - 2.69 $i$ & -0.18 + 0.05 $i$ & ---&  -2.69 - 2.28 $i$  &  -0.13 - 0.22 $i$\\
\midrule

& $H_{1,\frac{1}{2}} ~(V_{ub}V_{us}^*) $ &$H_{1,\frac{1}{2}} ~(V_{cb}V_{cs}^*)$&$H_{1,\frac{1}{2}} ~(V_{tb}V_{ts}^*)$ & $H_{0,\frac{1}{2}}~(V_{ub}V_{us}^*) $ &$H_{0,\frac{1}{2}}~(V_{cb}V_{cs}^*)$&$H_{0,\frac{1}{2}} ~(V_{tb}V_{ts}^*)$ \\
\midrule
$\mathcal{S}(\Lambda_b^0 \rightarrow pK^{*-})$& -10.36 &--- & - 0.23 &28.13 &---&0.62\\
$\mathcal{NC}(\Lambda_b^0 \rightarrow pK^{*-})$& -5.14 + 4.07 $i$ &--- & -0.16 + 0.13 $i$ & -3.14 + 16.76 $i$ &---&0.05 + 0.36 $i$\\
$\mathcal{C}(\Lambda_b^0 \rightarrow pK^{*-})$&---& 7.30 + 1.82 $i$ &0.14 - 0.05 $i$ &---&-2.34 + 2.49 $i$ & 0.02 - 0.03 $i$\\
\bottomrule
\bottomrule
\end{tabular}\label{Helicity Lb to pKstar}

\vspace{2cm}

\caption{Helicity amplitudes of $\Lambda_b^0\rightarrow p\rho^- (10^{-7})$}
\centering
\begin{tabular}{cccccccc}
\toprule
\toprule
 decay modes & $H_{0,-\frac{1}{2}}~(V_{ub}V_{ud}^*) $ &$H_{0,-\frac{1}{2}} ~(V_{cb}V_{cd}^*)$&$H_{0,-\frac{1}{2}} ~(V_{tb}V_{td}^*)$ & $H_{-1,-\frac{1}{2}} ~(V_{ub}V_{ud}^*) $ &$H_{-1,-\frac{1}{2}} ~(V_{cb}V_{cd}^*)$&$H_{-1,-\frac{1}{2}}~(V_{tb}V_{td}^*)$ \\
\midrule
$\mathcal{S}(\Lambda_b^0 \rightarrow p\rho^-  )$&141.80 & ---&3.14&28.02&---&0.62 \\
$\mathcal{NC}(\Lambda_b^0 \rightarrow p\rho^-  )$&-13.08 + 8.54 $i$ &---&-0.47 + 0.19 $i$&14.45 - 27.06 $i$ &---&0.01 - 0.46 $i$\\
$\mathcal{C}(\Lambda_b^0 \rightarrow p\rho^-  )$&---&-21.12 - 12.71 $i$&-0.37 - 0.09 $i$&---&- 9.45 - 4.37 $i$&-0.30 - 0.19 $i$\\
\midrule

& $H_{1,\frac{1}{2}} ~(V_{ub}V_{ud}^*) $ &$H_{1,\frac{1}{2}} ~(V_{cb}V_{cd}^*)$&$H_{1,\frac{1}{2}} ~(V_{tb}V_{td}^*)$ & $H_{0,\frac{1}{2}} ~(V_{ub}V_{ud}^*) $ &$H_{0,\frac{1}{2}} ~(V_{cb}V_{cd}^*)$&$H_{0,\frac{1}{2}} ~(V_{tb}V_{td}^*)$ \\
\midrule
$\mathcal{S}(\Lambda_b^0 \rightarrow p\rho^-  )$&-9.25&---&-0.20&
28.75&---&0.64\\
$\mathcal{NC}(\Lambda_b^0 \rightarrow p\rho^-  )$& - 12.31 + 6.38 $i$&---& - 0.10 - 0.79 $i$&2.22 + 13.81 $i$&---&0.32 + 0.31$ i$\\
$\mathcal{C}(\Lambda_b^0 \rightarrow p\rho^-  )$&---&11.81 + 6.52 $i$&0.31 + 0.17 $i$&---&-10.11 + 0.39 $i$&-0.11 - 0.08 $i$\\
\bottomrule
\bottomrule
\end{tabular}\label{Helicity Lb to prho}
\end{sidewaystable}

\newpage

\subsection{Discussions}
Based on above numerical results, some essential discussions are in order\textcolor{red}{:}  
\begin{itemize}
    \item First of all, one can see that the short-distance  amplitudes of tree and penguin are both of purely imaginary for $\Lambda^{0}_{b}\to p\pi^{-}$ and $pK^{-}$, but \deleted{ and} real for $\Lambda^{0}_{b}\to pK^{*-},p\rho^{-}$ and $\Lambda\phi$. It means that there is no relative strong phase to derive $CP$ asymmetry if only short distance contribution is considered. On the contrary, the long-distance ones investigated with final state re-scattering mechanism are generally complex, and an obvious strong phase source is provided.
    \item The long-distance charm triangle loop re-scattering contributions, as a component of non-factorizable penguin amplitudes, are found to be comparable and non-negligible relative to short-distance penguin amplitudes. This is particularly essential for the $\Lambda^{0}_{b}\to pK^{*-}, p\rho^{-}$ and $\Lambda \phi$ channels. As we will see that these contributions play an indispensable role in predicting the decay rate and triple product asymmetry of $\Lambda^{0}_{b}\to \Lambda\phi$. Both the charmless and charm triangle loop re-scattering amplitudes, which belong to $V_{tb}V^{*}_{tq}$ ($q=d,s$), are small and negligible due to the large suppression from the small Wilson coefficients $a_{4,6}$ relative to $a_{1}$ \cite{Lu:2009cm}. 
    \item It is stressed that the decays $\Lambda^{0}_{b}\to pK^{-}$ and $pK^{*-}$ are penguin dominant after incorporating the CKM enhancement factor, although the strong dynamics contribution involved in external $W$-emission $T$ topological graph is much larger than penguin diagram $P$. This can be easily seen from the ratio $\left|\frac{V_{tb}V^{*}_{ts}}{V_{ub}V^{*}_{us}}\right|\approx 50$, which indicates the penguin amplitude is enhanced almost $50$ times. Hence, the branching ratio of $\Lambda^{0}_{b}\to pK^{-}$ predicted from the calculation with only tree operators is one order smaller than the experimental measurement \cite{Zhu:2016bra,Wei:2009np}. The decay $\Lambda^{0}_{b}\to \Lambda\phi$ is also dominated by penguin ones like $\Lambda^{0}_{b}\to pK^{-}$, since the $W$-exchanged tree amplitude is largely CKM suppressed \cite{Rui:2022jff}. While it is different for $\Lambda^{0}_{b}\to p\pi^{-}$ and $p\rho^{-}$ where $\left|\frac{V_{tb}V^{*}_{td}}{V_{ub}V^{*}_{ud}}\right|\approx 2$, no remarkable enhancement emerges for penguin contribution hence the tree diagram is overly dominant.
    \item It is remarkably observed that the strong dynamics amplitude of $W$-exchange in $\Lambda^{0}_{b}\to \Lambda\phi$ is not suppressed compared to penguin ones. This is similar to $B$ meson decays where an obvious long-distance contribution to $W$-exchange is induced from final-state interactions, even if its short-distance amplitudes are vanishing, for example, $\bar{B}^{0}\to D^{0}\pi^{0}$ in \cite{Cheng:2004ru}. Additionally, the highlighted non-factorizable contribution of charm penguin amplitude demonstrates again that long-distance re-scatterings are important for processes without over-dominated $T$ diagram.
    
    \item The power counting rule based on the SCET analysis is numerically verified in our work. Specifically, the $T$ topological diagram is dominant due to short-distance contributions, while the other tree diagrams $C, C', E, B$ are mainly induced by long-distance amplitudes. In our results, we find that the charmless triangle loop contributions, which give rise to nonfactorizable long-distance tree amplitudes, are all one order of magnitude smaller than the short-distance ones in $\Lambda^{0}_{b}\to p\pi^{-}, p\rho^{-}, pK^{-}, pK^{*-}$ channels. This behavior is consistent with the expectations from the power counting rule \cite{Leibovich:2003tw,Mantry:2003uz}. Additionally, it is also reasonable that the total penguin amplitude $P$ is one order of magnitude smaller than $T$ for the aforementioned channels.
    
    \item The relative magnitudes of amplitudes with different helicity configurations of the external $W$-emission $T$ diagram can be determined by imposing the chiral property of weak interactions. For the simpler case of $\Lambda^{0}_{b}\to p\pi^{-}, pK^{-}$, where the final meson is pseudoscalar and thus trivial for helicity analysis, the final proton spin is preferably anti-parallel to its direction of motion due to the chiral current $(V-A)\times(V-A)$. It suggests that the helicity amplitude $H_{-\frac{1}{2}}$ should be over dominant in the short-distance contribution from the $T$ topology.
    An approximate estimation of the relative ratio $|H_{+\frac{1}{2}}|/|H_{-\frac{1}{2}}|$ can be derived within the naive picture of helicity flip, yielding a factor of $\Lambda_{\text{QCD}}/m_{b}$. This is indeed confirmed by comparing our results, as shown in Table~\ref{Helicity Lb to pK} and \ref{Helicity Lb to ppi}. The analogous intuition can be generalized to the decays $\Lambda^{0}_{b}\to p\rho^{-}, pK^{*-}$. For these decays, we conclude that the helicity amplitudes from the $T$-topological diagram are expected to satisfy
  \begin{equation}
 \begin{aligned}
 H_{0,-\frac{1}{2}}: H_{-1,-\frac{1}{2}}: H_{0,+\frac{1}{2}}:H_{+1,+\frac{1}{2}}\sim 1:\frac{\Lambda_{QCD}}{m_{b}}:\frac{\Lambda_{QCD}}{m_{b}}:\left(\frac{\Lambda_{QCD}}{m_{b}}\right)^{2} .
         \end{aligned}
\end{equation}
One can confirm that the above power relation is approximately consistent with our numerical results in the Table~\ref{Helicity Lb to pKstar} and  \ref{Helicity Lb to prho} with $\Lambda_{QCD}/{m_{b}} \sim \mathcal{O}(10^{-1})$.
\end{itemize}

\section{$\Lambda^{0}_{b}\to p\pi^{-}$ and $pK^{-}$ decays}\label{ppi-pk}

\subsection{Numerical results}
The branching ratios and direct CP asymmetries for $\Lambda^{0}_{b}\to p\pi^{-}$ and $pK^{-}$ are given as
\begin{equation}
 \begin{aligned}
BR\left[\Lambda^{0}_{b}\to p\pi^{-}(pK^{-})\right]=\frac{|p_{c}|}{8\pi M^{2}_{\Lambda_{b}}\Gamma_{\Lambda^{0}_{b}}}\frac{1}{2}\left(\left|H_{+1/2}\right|^{2}+\left|H_{-1/2}\right|^{2}\right),~~~~~a^{dir}_{CP}=\frac{\Gamma-\bar{\Gamma}}{\Gamma+\bar{\Gamma}} ,
         \end{aligned}
\end{equation}
where $p_{c}$ is the final proton momentum in the $\Lambda^{0}_{b}$ rest frame, and the factor $1/2$ in $BR\left[\Lambda^{0}_{b}\to p\pi^{-}(pK^{-})\right]$ accounts for the initial spin average. $H_{\pm \frac{1}{2}}$ are the helicity amplitudes listed in Table \ref{Helicity Lb to pK} and Table \ref{Helicity Lb to ppi}. As previously stressed, the branching ratios and direct CP asymmetry of decays  $\Lambda^{0}_{b}\to p\pi^{-}$ and $pK^{-}$ provide fruitful implications for controlling the final-state interactions of $\Lambda^{0}_{b}$ baryon decay. A global analysis is performed,  thus the model parameters $\Lambda_{\textbf{charm}}$ and $\Lambda_{\textbf{charmless}}$ are uniquely determined. Consequently, the final numerical results for these two decays are in close agreement with the experimental measurements.

Besides \deleted{for} the branching ratios and direct CP asymmetries, there are many other asymmetry parameters incorporating the interference terms of different partial waves that are of interest in baryon decays. These parameters are expected to reveal more about the helicity structure of the weak Hamiltonian, as discussed before, and are more sensitive to different strong dynamical approaches. Hence, they provide powerful tests on the theoretical side. For example, the asymmetry parameters measured in $\Lambda^{+}_{c}$ decay provide \deleted{eventually impose} an important test for non-perturbative methods \cite{BESIII:2023wrw}. Here, we extend similar study to   $\Lambda^{0}_{b}\to p\pi^{-}$ and $pK^{-}$   decays, with asymmetry parameters  defined as \cite{Chen:2019hqi}
\begin{equation}
 \begin{aligned}
\alpha=\frac{\left|H_{+1/2}\right|^{2}-\left|H_{-1/2}\right|^{2}}{\left|H_{+1/2}\right|^{2}+\left|H_{-1/2}\right|^{2}},~~~\beta=\frac{2\mathcal{I}m\left(H_{+1/2}H^{*}_{-1/2}\right)}{\left|H_{+1/2}\right|^{2}+\left|H_{-1/2}\right|^{2}},~~~\gamma=\frac{2\mathcal{R}e\left(H_{+1/2}H^{*}_{-1/2}\right)}{\left|H_{+1/2}\right|^{2}+\left|H_{-1/2}\right|^{2}} .
         \end{aligned}
\end{equation}
The corresponding parameters $\bar \alpha$, $\bar \beta$ and $\bar \gamma$ for the anti-baryon decays can also be defined similarly. Then, one can get the average asymmetry parameters and their associated CP asymmetries as \cite{Donoghue:1986hh}
\begin{equation}
 \begin{aligned}
\langle\alpha\rangle=\frac{\alpha-\bar{\alpha}}{2},~~\langle\beta\rangle=\frac{\beta-\bar{\beta}}{2},~~\langle\gamma\rangle=\frac{\gamma+\bar{\gamma}}{2},
         \end{aligned}
\end{equation}
\begin{equation}\label{AsymmetryparametersCPV1}
 \begin{aligned}
a^{\alpha}_{CP}=\frac{\alpha+\bar{\alpha}}{2},~~a^{\beta}_{CP}=\frac{\beta+\bar{\beta}}{2},~~a^{\gamma}_{CP}=\frac{\gamma-\bar{\gamma}}{2}
       .  \end{aligned}
\end{equation}
The relation between partial wave and helicity amplitudes are linear and trivial \cite{Chen:2019hqi}
\begin{equation}
	\begin{aligned}
    \mathcal{H}_{+{1\over2}}=\frac{1}{\sqrt{2}} (S+P),~~~~
    \mathcal{H}_{-{1\over2}}=\frac{1}{\sqrt{2}} (S-P) .
	\end{aligned}
\end{equation}

Next, one can define the CPV observables associated with each partial wave amplitude analogy to the direct CP asymmetry \cite{Roy:2019cky}
\begin{equation}
	\begin{aligned}
a^{S}_{CP}=\frac{|S|^{2}-|\bar{S}|^{2}}{|S|^{2}+|\bar{S}|^{2}},~~~~
a^{P}_{CP}=\frac{|P|^{2}-|\bar{P}|^{2}}{|P|^{2}+|\bar{P}|^{2}} .
	\end{aligned}
\end{equation}
The global direct CP asymmetry  is
\begin{equation}
	\begin{aligned}
a^{dir}_{CP}=\frac{\Gamma-\bar{\Gamma}}{\Gamma+\bar{\Gamma}}=\frac{|S|^{2}-|\bar{S}|^{2}+|P|^{2}-|\bar{P}|^{2}}{|S|^{2}+|\bar{S}|^{2}+|P|^{2}+|\bar{P}|^{2}} ,
	\end{aligned}
\end{equation}
which indicates that the global $CP$ asymmetry might be suppressed if the cancellation of $CP$ asymmetries between $a^{S}_{CP}$ and $a^{P}_{CP}$ arises. From Table~\ref{BR of ppi and pk}, it is easy to note that the partial wave $CP$ asymmetries of $a^{S}_{CP}$ and $a^{P}_{CP}$ are salient, while the global $CPV$ is small owing to the remarkable cancellation between them. It is very distinctive compared to mesonic decays due to the helicity property of baryons. This phenomenon is first discovered in the recent work \cite{Yu:2024cjd}, where a complete $PQCD$ calculation is performed on $\Lambda^{0}_{b}\to p\pi^{-}$ and $pK^{-}$. Here, this cancellation is confirmed again under the approach of FSIs with the hadronic loop method.
In  Table~\ref{BR of ppi and pk}, we list our results of BRs, direct $CPV$s, and partial wave $CP$ asymmetries with the model parameters $\Lambda_{\textbf{charmless}}=0.5\pm0.1$ and $\Lambda_{\textbf{charm}}=1.0\pm0.1$, comparing with  results from other works. 
In  Table~\ref{asymmetry 1}, we list numerical results of  asymmetry parameters $\langle\alpha\rangle,\langle\beta\rangle,\langle\gamma\rangle$ and their associated CPVs $a^{\alpha}_{CP},a^{\beta}_{CP},a^{\gamma}_{CP}$ for $\Lambda_b^0 \rightarrow p\pi^{-}$ and $pK^{-}$ decays. 

\begin{table}[htbp]
\caption{BRs, Direct $CPA$s, $S$- and $P$-wave $CPA$s of $\Lambda_b^0 \rightarrow p\pi^{-}$ and $pK^{-}$ decays calculated in this work comparing with  other works. }
\centering
\begin{tabular}{cccccccccc}
\toprule
\toprule
 & BR($10^{-6}$)& $\alpha$ & Direct CP($10^{-2}$) &$a^{S}_{CP}$ &$a^{P}_{CP}$  \\
\midrule
$\Lambda_b^0 \rightarrow pK^-$ \\
FSI (This work)    
& $4.98^{+3.61}_{-1.98}$ 
& $0.97^{+0.02}_{-0.05}$
&${-9}_{-2}^{+2}$
&$0.12_{-0.04}^{+0.03}$
&$-0.24_{-0.09}^{+0.07}$
\\
PQCD \cite{Yu:2024cjd}
& $2.9$ 
& $0.38$
& $-5.8$
& $-0.05$
& $-0.23$
\\
QCDF \cite{Zhu:2018jet}
& $2.17^{+0.98+0.60+0.33}_{-0.47-0.58-0.23}$ 
& $0.27^{+0.19}_{-0.14}$
& $10$
&---
&---
\\
Bag model \cite{Geng:2020ofy}
& $6.0$ 
& $0.297$
& $-19.6$
&---
&---
\\
GFA \cite{Hsiao:2017tif}
& $4.49^{+0.84}_{-0.39}\pm0.26\pm0.59$ 
&---
& ${6.7}^{+0.3}_{-0.2}\pm0.3$
&---
&---
\\
Exp \cite{LHCb:2024iis}
& $5.5\pm1.0$
&---
& $-1.1\pm0.7\pm0.4$
&---
&---
\\
\toprule
$\Lambda_b^0 \rightarrow p\pi^-$ \\
FSI (This work)  
&$4.28_{-0.30}^{+0.66}$
&$-0.75^{+0.18}_{-0.13}$
&${-2} _{-5}^{+6}$
&$-0.22_{-0.11}^{+0.15}$
&$0.51_{-0.25}^{+0.25}$
\\
PQCD \cite{Yu:2024cjd}
& $3.3$ 
& $-0.81$
& $4.1$
& $0.15$
& $-0.07$
\\
QCDF \cite{Zhu:2018jet}
& $4.30^{+0.27+1.18+0.69}_{-0.19-1.16-0.45}$ 
& $-0.98^{+0.00}_{-0.01}$
& $-0.337$
&---
&---
\\
Bag model \cite{Geng:2020ofy}
& $5.0$ 
& $-0.856$
& $1.4$
&---
&---
\\
GFA \cite{Hsiao:2017tif}
& $4.25$ 
&---
& $-3.9\pm0.4$
&---
&---
\\
Exp \cite{LHCb:2024iis}
& $4.6\pm0.8$ 
&---
& $0.2\pm 0.8\pm0.4$
&---
&---
\\
\bottomrule
\bottomrule
\end{tabular}\label{BR of ppi and pk}
\end{table}

\begin{table}[htbp]
\caption{Average asymmetry parameters and their $CPV$s for $\Lambda_b^0\rightarrow p\pi^-$ and $ pK^-$ decays. }
\centering
\begin{tabular}{cccccccc}
\toprule
\toprule
  & $\langle\alpha\rangle$ &$a^{\alpha}_{CP}$&$ \langle\beta\rangle$ & $a^{\beta}_{CP}$ &$\langle\gamma\rangle$&$a^{\gamma}_{CP}$ \\
\midrule
$\Lambda_b^0 \rightarrow pK^-$& $0.20^{+0.02}_{-0.05}$ & $0.77^{+0.05}_{-0.07}$ & $-0.25^{+0.08}_{-0.10}$ &  $0.50^{+0.05}_{-0.11}$    & $0.18^{+0.04}_{-0.06}$  & $-0.15^{+0.28}_{-0.21}$  \\
$\Lambda_b^0 \rightarrow p\pi^-$&  $-0.08^{+0.02}_{-0.01}$ & $-0.67^{+0.20}_{-0.15}$  &$0.49^{+0.14}_{-0.11}$ & $0.15^{+0.14}_{-0.17}$  &$-0.29^{+0.14}_{-0.11}$ & $0.44^{+0.05}_{-0.14}$ &\\
\midrule
\midrule
\end{tabular}\label{asymmetry 1}
\end{table}


\newpage
\subsection{Discussions}

Some discussions are in order:

\begin{itemize}
    \item The regulated parameter of charm hadronic loops $\Lambda_{\textbf{charm}} = 1.0$ is obviously larger than that of charmless loops $\Lambda_{\textbf{charmless}} = 0.5$. One might understand this qualitatively by observing that the form factor $\mathcal{F}(\Lambda, m_{k})$ defined in eq.(\ref{FF})  approaches $1$ as $\Lambda \to \infty$, indicating that the particles involved in rescatterings are completely point-like under this limit. Taking $\Lambda^{0}_{b}\to p\pi^{-}$ as an example, the residual energy via $\Lambda^{0}_{b}\to \Lambda^{+}_{c}D^{(*)}\to p\pi^{-}$ is significantly lower than that via $\Lambda^{0}_{b} \to p\pi^{-}(p\rho^{-}) \to p\pi^{-}$. Consequently, the latter process proceeds due to higher energies, implying that the QCD substructure of proton and $\pi/\rho$ mesons are more considerable. Hence, the charmless loop parameter $\Lambda_{\textbf{charmless}}$ is expected to be smaller.
    
  \item The model parameters $\Lambda_{\textbf{charm}}$ and $\Lambda_{\textbf{charmless}}$ determined in this work are expected to be applicable to similar rescattering triangle loops in other channels of $\Lambda^{0}_{b}$ and even other $b$-baryon charmless hadronic decays. This expectation is based on the approximate SU(3) flavor symmetry for light hadron groups and heavy-quark symmetry for heavy baryons. For example, it is anticipated that these parameters will show similar capability in calculating decays such as $\Lambda^{0}_{b} \to p a_{1}$, $pK_{1}$, $pf_{0}(980)$, $\Lambda(1520)\phi$, and potentially many more decays within this framework in the future.
  
  \item The dependence of $\Lambda^{0}_{b}\to p\pi^{-}$ and $\Lambda^{0}_{b}\to pK^{-}$ branching ratios on the parameters $\Lambda_{\textbf{charm}}$ and $\Lambda_{\textbf{charmless}}$ can be well understood by recognizing that these decays are dominated by tree and penguin operators, respectively. As previously emphasized, $\Lambda^{0}_{b}\to p\pi^{-}$ is primarily driven by external $W$-emission amplitudes, which are short-distance interactions and thus not sensitive to the model parameter $\Lambda_{\textbf{charm}}$, but is slightly sensitive to $\Lambda_{\textbf{charmless}}$. Conversely, the $\Lambda^{0}_{b}\to pK^{-}$ is dominated by penguin amplitudes due to CKM enhancement factor. The associated long distance charm hadron loop contribution, which depends quartically on model parameter $\Lambda_\textbf{charm}$,  is the non-factorizable part of charm penguin amplitudes. As a result, its branching ratio depends on $\Lambda_\textbf{charm}^8$ when ignoring the tree and short distance penguin contributions, leading to a significant variation with the $\Lambda_\textbf{charm}$. Finally, the $BR(\Lambda_b^0 \to p K^-)$ suffers from large uncertainties, however its variation does not strictly follow $\Lambda_\textbf{charm}^8$ power rule with $1 \pm 0.1 $ since the short distance penguin amplitude is also comparable and does not depend on model parameter $\Lambda_\textbf{charm}$. 
  As mentioned before, direct CP asymmetries are expected to be insensitive to model parameters because the dependence on the parameters is largely canceled out in the ratios, thereby reducing the theoretical uncertainties on these observables \cite{Jia:2024pyb}.
  \item The strong couplings used to describe the effective hadronic interactions in re-scatterings suffer significant uncertainties and can vary widely in different references \cite{Aliev:2010yx,Aliev:2010nh,Azizi:2014bua,Aliev:2006xr,Aliev:2009ei,Yu:2016pyo,Janssen:1996kx,
   Azizi:2015tya,Ballon-Bayona:2017bwk,Jiang:2024equ,Jin:2024zyy,
   Casalbuoni:1996pg,Ronchen:2012eg}. While we have chosen values based on the extractions from experimental data and LCSRs calculations, improving the precision of these couplings is essential for better  theoretical precisions, thereby advancing our understanding of the dynamics of $b$-baryon decays in the future.  
  \item The comparison of our results with those obtained from different theoretical methods is presented in Table \ref{BR of ppi and pk}. The branching ratios calculated by various approaches show good agreement, while our prediction for the asymmetry parameter $\alpha$ in the $pK^{-}$ channel stands out as notably different, thereby offering a distinct test for final-state interactions (FSIs) in future experiments. The partial wave CP asymmetries in our study also differ from those predicted by the PQCD approach. In PQCD, the CP asymmetry in $\Lambda^{0}_{b}\to p\pi^{-}$ is small due to the cancellation between the $a^{S}_{CP}$ and $a^{P}_{CP}$ terms, while the CPA in $\Lambda^{0}_{b}\to pK^{-}$ is small since it is dominated by the $S$-wave contribution. In our analysis, both decays exhibit CPAs that result from the cancellations between $a^{S}_{CP}$ and $a^{P}_{CP}$. 
Furthermore, the average asymmetry parameters $\langle\beta\rangle$ and $\langle\gamma\rangle$, along with their associated CP asymmetries, are listed in Table \ref{asymmetry 1} for the first time.
  
  \item The cancellation between $a^{S}_{CP}$ and $a^{P}_{CP}$ can be further confirmed by examining Table \ref{helicityCPV1}, where the CP asymmetries from each helicity amplitude are presented. It is evident that the CP violation in $\Lambda^{0}_{b}\to p\pi^{-}$ is expected to be small due to the dominance of $H_{-\frac{1}{2}}$, while the CP violation in $\Lambda^{0}_{b}\to pK^{-}$ is also anticipated to be small as a result of the dominance of $H_{+\frac{1}{2}}$ that can be verified by analyzing the asymmetry parameter $\alpha(\Lambda^{0}_{b}\to pK^{-})$.

\begin{table}[htbp]
\caption{$CP$ violation of helicity amplitude of $\Lambda_b^0 \rightarrow p\pi^{-}$ and $pK^{-}$ decays. }
\centering
\begin{tabular}{cccccccccc}
\toprule
\toprule
Decay modes & CPV($H_{-\frac{1}{2}}$) & CPV($H_{\frac{1}{2}}$) &  \\
\midrule
$\Lambda_b^0 \rightarrow pK^-$     
& $-0.88^{+0.13}_{-0.07}$ 
&$0.02^{+0.02}_{-0.04}$ 
\\
$\Lambda_b^0 \rightarrow p\pi^-$
&$0.03_{-0.03}^{+0.05}$
&$-0.27_{-0.05}^{+0.10}$
\\
\bottomrule
\bottomrule
\end{tabular}\label{helicityCPV1}
\end{table}

  \item The asymmetry parameter $\alpha$ for $\Lambda^{0}_{b}\to p\pi^{-}$ is approaching to $-1$ under the heavy quark symmetry and $(V-A)$ weak current interaction \cite{Manohar:2000dt}. The result obtained in our work, $-0.75$, is in   agreement with this leading order HQET prediction. However, there is still a 20\% deviation, which can be attributed to the power correction of heavy quark expansion. In our work, we actually consider the power-suppressed effects of $C,E,P...$ diagrams by estimating FSIs contributions. We hope a more comprehensive understanding of the bottom baryon charmless non-leptonic dynamics can be achieved. 
\end{itemize}

\section{$\Lambda^{0}_{b}\to p\rho^{-},pK^{*-}$ and $\Lambda\phi$ decays}\label{prhoklambdaphi}

\subsection{Numerical results}

Next, we will explore the decay mode of $\Lambda^{0}_{b}$ to a light baryon $B$ and a vector meson $V$, including the three channels $\Lambda^{0}_{b}\to p\rho^{-}$, $pK^{*-}$, and $\Lambda\phi$.
The branching ratios for these associated decays are defined as
\begin{equation}
	\begin{aligned}
BR\left[\Lambda^{0}_{b}\to BV\right]=\frac{|p_{c}|}{8\pi M^{2}_{\Lambda_{b}}\Gamma_{\Lambda^{0}_{b}}}\frac{1}{2}\left(\left|H_{0,+1/2}\right|^{2}+\left|H_{0,-1/2}\right|^{2}+\left|H_{+1,+1/2}\right|^{2}+\left|H_{-1,-1/2}\right|^{2}\right) ,
	\end{aligned}
\end{equation}
where four independent helicity amplitudes are involved in the $\Lambda^{0}_{b}\to BV$ channels. The decay asymmetry parameters are \cite{Hong:2022prk} 
\begin{equation}
	\begin{aligned}
\alpha^{\prime}=\frac{\left|H_{+1,+\frac{1}{2}}\right|^{2}-\left|H_{-1,-\frac{1}{2}}\right|^{2}}{\left|H_{+1,+\frac{1}{2}}\right|^{2}+\left|H_{-1,-\frac{1}{2}}\right|^{2}},~~\beta^{\prime}=\frac{\left|H_{0,+\frac{1}{2}}\right|^{2}-\left|H_{0,-\frac{1}{2}}\right|^{2}}{\left|H_{0,+\frac{1}{2}}\right|^{2}+\left|H_{0,-\frac{1}{2}}\right|^{2}},~~\gamma^{\prime}=\frac{\left|H_{+1,+\frac{1}{2}}\right|^{2}+\left|H_{-1,-\frac{1}{2}}\right|^{2}}{\left|H_{0,+\frac{1}{2}}\right|^{2}+\left|H_{0,-\frac{1}{2}}\right|^{2}} ,
	\end{aligned}
\end{equation}
and longitudinal polarization of final baryon $B$
\begin{equation}
	\begin{aligned}
P_{L}=\frac{\left|H_{+1,+\frac{1}{2}}\right|^{2}-\left|H_{-1,-\frac{1}{2}}\right|^{2}+\left|H_{0,+\frac{1}{2}}\right|^{2}-\left|H_{0,-\frac{1}{2}}\right|^{2}}{\left|H_{+1,+\frac{1}{2}}\right|^{2}+\left|H_{-1,-\frac{1}{2}}\right|^{2}+\left|H_{0,+\frac{1}{2}}\right|^{2}+\left|H_{0,-\frac{1}{2}}\right|^{2}} .
	\end{aligned}
\end{equation}
These parameters are not all independent, and they are related to each other as
\begin{equation}
	\begin{aligned}
P_{L}=\frac{\beta+\alpha\cdot\gamma}{1+\gamma} .
	\end{aligned}
\end{equation}

These asymmetry parameters could be extracted through complete polarized angular analysis as done for $\Lambda^{+}_{c}\to p\phi$ decays in Ref. \cite{Hong:2022prk}, or partial wave analysis as for $\Lambda^{+}_{c}\to \Lambda\rho^{+}$ \cite{BESIII:2022udq}. The corresponding average asymmetry parameters and their $CP$ asymmetries will be defined by taking the difference and summation of these parameters and their $CP$ conjugates as
\begin{equation}
	\begin{aligned}
\langle\alpha^{\prime}\rangle=\frac{\alpha^{\prime}-\bar{\alpha}^{\prime}}{2},~~\langle\beta^{\prime}\rangle=\frac{\beta^{\prime}-\bar{\beta}^{\prime}}{2},~~\langle\gamma^{\prime}\rangle=\frac{\gamma^{\prime}+\bar{\gamma}^{\prime}}{2},~~\langle{P_{L}}\rangle=\frac{P_{L}-\bar{P}_{L}}{2}
	\end{aligned}
\end{equation}
\begin{equation}
	\begin{aligned}
a_{CP}^{\alpha^{\prime}}=\frac{\alpha^{\prime}+\bar{\alpha}^{\prime}}{2},~~a_{CP}^{\beta^{\prime}}=\frac{\beta^{\prime}+\bar{\beta}^{\prime}}{2},~~a_{CP}^{\gamma^{\prime}}=\frac{\gamma^{\prime}-\bar{\gamma}^{\prime}}{2},~~a_{CP}^{{P_{L}}}=\frac{P_{L}+\bar{P}_{L}}{2}
.	\end{aligned}\label{AsymmetryparametersCPV2}
\end{equation}
For the decay $\Lambda_b^0 \rightarrow \Lambda \phi$, additional observables known as $T$-odd triple product asymmetries ($TPA$s) can be involved if considering the secondary decays $\phi\to K^{+}K^{-}$ and $\Lambda\to p\pi^{-}$, as this introduces more angular variables. The specific definitions based on the helicity formalism and associated complete angular distribution function can be found in Ref. \cite{Rui:2022jff,Geng:2021sxe}. In our analysis, we provide numerical predictions for these asymmetries within the approach of the final-state re-scattering.
\begin{align*}
A_T^1 &=-\frac{\alpha_{\Lambda}}{\sqrt{2}} \frac{\mathcal{I}m\left[H_{0 \frac{1}{2} } H_{-1 -\frac{1}{2} }^*+H_{0 -\frac{1}{2} } H_{1 \frac{1}{2} }^*\right]}{H_N}, \quad\quad
A_T^2  =-\frac{P_b}{\sqrt{2}} \frac{\mathcal{I}m\left[H_{-1 -\frac{1}{2}} H_{0 -\frac{1}{2} }^*+H_{1 \frac{1}{2} } H_{0 \frac{1}{2} }^*\right]}{H_N}, \\
A_T^3 &=\frac{P_b \alpha_{\Lambda}}{2 \sqrt{2}} \frac{\mathcal{I}m\left[H_{-1 -\frac{1}{2}} H_{0 \frac{1}{2} }^*-H_{1 \frac{1}{2} } H_{0 -\frac{1}{2} }^*\right]}{H_N}, \quad
A_T^4 =\frac{P_b \alpha_{\Lambda}}{2 \sqrt{2}} \frac{\mathcal{I}m\left[H_{-1 -\frac{1}{2} } H_{0 -\frac{1}{2} }-H_{ 1 \frac{1}{2}} H_{ 0 \frac{1}{2}}^*\right]}{H_N},\\
A_T^5  &=-\frac{P_b \pi \alpha_{\Lambda}}{4} \frac{\mathcal{I}m\left[H_{0 -\frac{1}{2} } H_{0 \frac{1}{2} }^*\right]}{H_N}, \quad\quad\quad\quad\quad
A_T^6 =-\frac{P_b \pi \alpha_{\Lambda}}{4} \frac{\mathcal{I}m\left[H_{1 \frac{1}{2} } H_{-1 -\frac{1}{2}}^*\right]}{H_N} ,
\end{align*}
where $H_{N}$ is a normalization factor
\begin{equation}
	\begin{aligned}
H_N=\left|H_{1 \frac{1}{2}}\right|^2+\left|H_{-1 -\frac{1}{2}}\right|^2+\left|H_{0 \frac{1}{2}}\right|^2+\left|H_{0 -\frac{1}{2}}\right|^2 
	. \end{aligned}
\end{equation}

One can define the true $T-$odd $CP$ asymmetries by taking off the pollution from strong interactions
\begin{equation}
	\begin{aligned}
a_{T,CP}^{i}=\frac{A_T^i-\bar{A}_T^i}{2}
,	\end{aligned}
\end{equation}
where the quantities $\bar{A}_T^i$ with $i= 1,...6$ correspond to the charge conjugates of the triple products.  Utilizing the latest experimental data for the asymmetry parameters $\alpha_{\Lambda}= 0.732 \pm 0.014$ and $\alpha_{\bar{\Lambda}}=-0.758 \pm 0.012$ associated with $\Lambda$ and $\bar{\Lambda}$ decays \cite{BESIII:2018cnd}, we can analyze these triple product asymmetries (TPAs). It is important to note that some of these observables are influenced by the initial polarization of $\Lambda^{0}_{b}$, denoted by $P_{b}$, which has not yet been firmly established by experimental data \cite{LHCb:2013hzx,CMS:2018wjk,LHCb:2020iux,Zhang:2023nip}. As an illustrative example, we consider $P_{b}=0.1$ in our numerical predictions for these observables, as discussed in \cite{Rui:2022jff}. Numerical results of BRs, Direct $CP$As and asymmetric parameters in $\Lambda^{0}_{b}\to p\rho^{-}$, $pK^{*-}$ and $\Lambda\phi$ decays and comparison with other approaches are summarized in   Table~\ref{BRs2}. The numerical results for   average asymmetry parameters $\langle\alpha^{\prime}\rangle,\langle\beta^{\prime}\rangle,\langle\gamma^{\prime}\rangle,\langle P_{L}\rangle$ and their associated CPAs for three channels  are summarized in   Table~\ref{CPAs2}.  Numerical results for triple products and its asymmetries in $\Lambda_b^0 \rightarrow \Lambda \phi$ decay calculated in FSIs and PQCD approach \cite{Rui:2022jff} are presented in Table~\ref{TPAs}.

\begin{table}[htbp]
\caption{Numerical results of BRs, Direct $CP$As and asymmetric parameters in $\Lambda^{0}_{b}\to p\rho^{-}$, $pK^{*-}$ and $\Lambda\phi$ decays and comparison with other approaches. }
\centering
\begin{tabular}{cccccccccc}
\toprule
\toprule
 & BR($10^{-5}$)& Direct $CP$ & $\alpha$& $\beta$& $\gamma$&$P_L$ \\
\midrule
$\Lambda_b^0 \rightarrow pK^{*-}$\\
FSI (this work)
&$1.35_{-0.71}^{+1.36}$
&$0.02_{-0.04}^{+0.04}$
&$0.60_{-0.06}^{+0.03}$
&$-0.85_{-0.05}^{+0.03}$
&$0.55_{-0.12}^{+0.11}$
&$-0.34_{-0.10}^{+0.08}$\\
PQCD \cite{Yu:2024cjd} 
&$0.302$
&$0.057$
&$-0.999$
&$-0.92$
&$0.11$
&---\\
QCDF \cite{Zhu:2018jet}
&$0.101$
&$0.311$
&---
&---
&---
&$-0.79$\\
GFA \cite{Hsiao:2017tif}
&$0.286$
&$0.197$
&---
&---
&---
&---\\
\midrule
$\Lambda_b^0 \rightarrow p\rho^-$ \\
FSI (this work)
&$1.34_{-0.19}^{+0.53}$
&$-0.24_{-0.03}^{+0.07}$
&$-0.26_{-0.44}^{+0.37}$
&$-0.71_{-0.10}^{+0.14}$
&$0.12_{-0.04}^{+0.15}$
&$-0.66_{-0.11}^{+0.13}$ 
\\
PQCD \cite{Yu:2024cjd} 
&$1.513$
&$-0.020$
&$-0.71$
&$-0.98$
&$0.04$
&---
\\
QCDF \cite{Zhu:2018jet}
&$0.747$
&$-0.319$
&---
&---
&---
&$-0.81$\\
GFA \cite{Hsiao:2017tif}
&$1.1$
&$-0.038$
&---
&---
&---
&---\\
\midrule
$\Lambda_b^0 \rightarrow \Lambda \phi$ \\
FSI (this work)
&$0.31_{-0.19}^{+0.43}$
&$-0.005_{-0.01}^{+0.02}$
&$0.72_{-0.06}^{+0.03}$
&$-0.61_{-0.13}^{+0.43}$
&$3.10_{-1.1}^{+4.2}$
&$0.39_{-0.18}^{+0.17}$
\\
PQCD \cite{Rui:2022jff}
&$0.69$
&$-0.01$
&---
&$-0.71$
&---
&$-0.79$\\
QCDF \cite{Zhu:2018jet}
&$0.0633$
&$0.016$
&---
&---
&---
&$-0.80$\\
GFA \cite{Hsiao:2017tif}
&$0.177$
&$0.014$
&---
&---
&---
&---\\
\bottomrule
\bottomrule
\end{tabular}\label{BRs2}
\end{table}

\begin{table}
\caption{Average asymmetry parameters and their $CPV$ on $\Lambda_b^0\rightarrow pK^{*-}, p\rho^- $ and $\Lambda \phi $ decays.}
\centering
\resizebox{\textwidth}{!}
{
\begin{tabular}{cccccccccc}
\toprule
\toprule
 & $\langle\alpha^{\prime}\rangle$ &$a^{\alpha^{\prime}}_{CP}$&$\langle\beta^{\prime}\rangle$ & $a^{\beta^{\prime}}_{CP}$ &$\langle\gamma^{\prime}\rangle$&$a^{\gamma^{\prime}}_{CP}$ &$\langle P_{L}\rangle$ & $a^{P_{L}}_{CP}$& \\
\midrule
$\Lambda_b^0 \rightarrow pK^{*-}$& $-0.04^{+0.02}_{-0.03}$ & $0.65^{+0.01}_{-0.04}$  &  $-0.06^{+0.02}_{-0.03}$ & $-0.80^{+0.01}_{-0.02}$   & $0.03^{+0.08}_{-0.03}$ & $0.52^{+0.08}_{-0.09}$  &$-0.31^{+0.05}_{-0.07}$ & $-0.03^{+0.03}_{-0.03}$ & 
\\
$\Lambda_b^0 \rightarrow p\rho^-$ &  $0.01^{+0.10}_{-0.13}$ & $-0.28^{+0.27}_{-0.31}$  &$0.09^{+0.07}_{-0.05}$ & $-0.81^{+0.07}_{-0.05}$ &$-0.07^{+0.05}_{-0.11}$& $0.20^{+0.22}_{-0.08}$  &$-0.72^{+0.11}_{-0.09}$ & $0.05^{+0.03}_{-0.02}$&
\\
$\Lambda_b^0 \rightarrow \Lambda \phi$& $0.02^{+0.04}_{-0.01}$& $0.70^{+0.04}_{-0.10}$&$0.01^{+0.05}_{-0.01}$ &$-0.62^{+0.37}_{-0.13}$&$0.38^{+1.70}_{-0.31}$&$2.72^{+2.53}_{-0.80}$&$0.34^{+0.17}_{-0.15}$&$0.05^{+0.06}_{-0.04}$&
\\
\midrule
\midrule
\end{tabular}\label{CPAs2}
}
\end{table}


\begin{table}
\caption{$CP$ violation of each helicity amplitude for $\Lambda_b\rightarrow pK^{*-}, p\rho^- $ and $\Lambda \phi $ decays. }
\centering
\begin{tabular}{cccccccccc}
\toprule
\toprule
Decay modes & $CPV$($H_{0,-\frac{1}{2}}$) & CPV($H_{-1,-\frac{1}{2}}$) & CPV($H_{1,\frac{1}{2}}$) & CPV($H_{0,\frac{1}{2}}$) &  \\
\midrule
$\Lambda_b^0 \rightarrow pK^{*-}$     
& $0.03^{+0.06}_{-0.08}$ 
&$0.18^{+0.09}_{-0.07}$ 
&$0.03^{+0.04}_{-0.02}$
&$-0.28^{+0.12}_{-0.28}$
\\
$\Lambda_b^0 \rightarrow p\rho^-$
&$-0.23_{-0.06}^{+0.09}$
&$-0.53_{-0.07}^{+0.18}$
&$-0.51_{-0.13}^{+0.24}$
&$0.33_{-0.17}^{+0.20}$
\\
$\Lambda_b^0 \rightarrow \Lambda \phi$
&$-0.11_{-0.25}^{+0.08}$
&$-0.04_{-0.02}^{+0.03}$
&$0.05_{-0.03}^{+0.08}$
&$-0.09_{-0.15}^{+0.06}$
\\
\bottomrule
\bottomrule
\end{tabular}\label{CPAs22}
\end{table}

\begin{table}
\caption{Triple products and its asymmetries in $\Lambda_b^0 \rightarrow \Lambda \phi$ decay calculated in FSIs and PQCD approach\cite{Rui:2022jff}. }
\centering
\begin{tabular}{cccccc}
\toprule
\toprule
  & $A_T^i$ & $\bar{A}_T^i $& $a_{T,CP}^{i}$ \\
\midrule
$ i=1 $ \\
FSI (This work) & $1.1^{+0.3}_{-0.2} \times 10^{-1} $ & $1.2^{+0.5}_{-0.3} \times 10^{-1} $ &$ -6.2^{+4.6}_{-17.9} \times 10^{-3} $ \\
PQCD & $-1.4 \times 10^{-2} $ & $1.3\times 10^{-2} $ &$ -1.4 \times 10^{-2} $ \\
\midrule
$ i=2 $\\
FSI (This work) &  $7.0_{-2.8}^{+4.6} \times 10^{-3} $ & $5.8_{-3.2}^{+5.1} \times 10^{-3}$ & $6.2^{+20.4}_{-4.8} \times 10^{-4} $ \\
PQCD  & $-6.9 \times 10^{-3} $ & $3.5\times 10^{-3} $ &$ -1.7 \times 10^{-3} $ \\
\midrule
$i=3 $ \\
FSI (This work) & $ -2.4_{-2.2}^{+1.4} \times 10^{-3} $ & $ -2.2_{-1.7}^{+1.2} \times 10^{-3} $&$ -6.8^{+6.9}_{-35.7} \times 10^{-5}$\\
PQCD  & $-1.8 \times 10^{-3} $ & $-5.7\times 10^{-4} $ &$ -0.6 \times 10^{-3} $ \\
\midrule
$i=4 $\\
FSI (This work)& $6.1_{-0.6}^{+0.5} \times 10^{-3} $ &  $6.3_{-0.7}^{+0.7} \times 10^{-3}$ &$-5.4_{-38.8}^{+5.1} \times 10^{-5}$\\
PQCD  & $2.8 \times 10^{-3} $ & $1.8\times 10^{-3} $ &$ 0.5 \times 10^{-3} $ \\
\midrule
$i=5$\\
FSI (This work)&$ -5.5_{-0.9}^{+3.3} \times 10^{-3} $ & $ -6.7_{-0.6}^{+1.5} \times 10^{-3} $ & $6.0_{-4.7}^{+15.3} \times 10^{-4}$\\
PQCD  & $2.4 \times 10^{-3} $ & $-3.6\times 10^{-3} $ &$ 3.0 \times 10^{-3} $ \\
\midrule
$i=6 $\\
FSI (This work)& $ -1.2_{-0.7}^{+0.4} \times 10^{-2}$ & $ -1.2_{-0.6}^{+0.3} \times 10^{-2} $ & $ -7.7_{-23.5}^{+9.7} \times 10^{-5}$\\
PQCD  & $-5.9 \times 10^{-4} $ & $-5.5\times 10^{-4} $ &$ -0.2 \times 10^{-4} $ \\
\bottomrule
\bottomrule
\end{tabular}\label{TPAs}
\end{table}


\newpage

\subsection{Discussions}

Some   discussions are in order:
\begin{itemize}
    \item Our prediction for the decay rate of $\Lambda^{0}_{b}\to pK^{*-}$ closely matches the experimental measurement of $BR(\Lambda^{0}_{b}\to p\bar{K}^{0}\pi^{-})=(1.3\pm0.4)\times10^{-5}$ \cite{LHCb:2014yin}, indicating that the $pK^{*-}$ channel might be the most dominant subprocess in the three-body decay $\Lambda^{0}_{b}\to p\bar{K}^{0}\pi^{-}$. This dominance can be tested in future experiments. The branching ratios of $\Lambda^{0}_{b}\to p\rho^{-}$ and $\Lambda\phi$ are consistent with other theoretical predictions, except for that based on the QCDF approach with   diquark hypothesis. Under the diquark picture, some non-factorizable hard spectator contributions are missing,  hence it’s reasonability requires more experimental tests. In the Generalized Factorization Approach (GFA), the branching ratio of $\Lambda^{0}_{b}\to \Lambda\phi$ is enhanced by introducing an effective color number $N_{\textbf{eff}}=2$ to account for non-factorizable amplitudes, implying again that non-factorizable contributions may play a crucial role in decays without  the dominant tree amplitude.
    
    \item The direct CP asymmetry of $\Lambda^{0}_{b}\to pK^{*-}$ is anticipated to be significant in the QCDF and GFA, while it is expected to be small in PQCD and our current work. Therefore, a precise measurement of this asymmetry is crucial for a test in future experiments. Furthermore, the CP asymmetries stemming from each asymmetry parameter and helicity amplitude are provided in PQCD and our work. This information is essential not only for a more comprehensive dynamical analysis but also for potential experimental investigations, particularly if partial wave analysis is realized in future experiments. In the case of $\Lambda^{0}_{b}\to p\rho^{-}$, a large CP violation is predicted in our work and in the QCDF approach. This prediction could help experimenters to    target for  the search of CP violation in baryon decays, although the $\rho^-$ decay product $\pi^-\pi^0$ may not be good for a hadron experiment, like LHCb. It is worth noting that the $\Lambda^{0}_{b}\to p\rho^{-}$ decay channel is predominantly governed by the helicity amplitude $H_{0,-\frac{1}{2}}$, implying that the total CP violation in $p\rho^{-}$ is nearly equivalent to that of $H_{0,-\frac{1}{2}}$ by comparing to Table \ref{CPAs22}. On the other hand, the CP violation in the $\Lambda^{0}_{b}\to \Lambda\phi$ decay, shown in Table \ref{BRs2}, is expected to be very small as it is dominated  by penguin contribution. From Table \ref{Lb to Lambdaphi}, one can see that the tree contribution for this decay is suppressed by CKM matrix elements, while  the large helicity amplitudes listed there, are proportional to $V_{cb}V_{cs}^*$ or  $V_{tb}V_{ts}^*$. without weak phase.
    
    \item Asymmetry parameters, as previously emphasized, are highly sensitive to the phases of amplitudes, making them valuable indicators for testing various dynamical methods and models. In our analysis, we provide predictions for four parameters: $\alpha^{\prime},\beta^{\prime},\gamma^{\prime}$, and $P_{L}$. It is evident that these parameters exhibit variations across different dynamical approaches, highlighting the importance of experimental measurements to discern between the predictions. For the longitudinal polarization parameter $P_{L}$ in the $p\rho^{-}$ decay, it is expected to approach $-1$ based on heavy quark symmetry and the chiral properties of the charged weak current \cite{Manohar:2000dt}. This expectation is consistent with our results. However, in the case of $\Lambda\phi$ decay, our prediction for $P_{L}$ differs in sign from the predictions of PQCD and the QCDF. This discrepancy requires further experimental investigations to clarify the true nature of these asymmetry parameters and their implications for the underlying dynamics of the decays.  
    
    \item The CP asymmetries defined in Eqs.(\ref{AsymmetryparametersCPV1},\ref{AsymmetryparametersCPV2}) are indeed more robust and preferable compared to a direct extension like $\frac{\alpha+\bar{\alpha}}{\alpha-\bar{\alpha}}$, $\frac{\beta+\bar{\beta}}{\beta-\bar{\beta}},\frac{\gamma+\bar{\gamma}}{\gamma-\bar{\gamma}}$ since they are already dimensionless. Moreover, there is no inherent principle that ensures these definitions yield numerical results within the range of $-1$ to 1. It is also worth considering that the direct extension definitions may introduce large uncertainties when the denominators are very small. Hence, we adopt the definitions in Eqs.(\ref{AsymmetryparametersCPV1},\ref{AsymmetryparametersCPV2}) in the current work.
    
    \item The triple product asymmetry parameter is a scalar quantity that is defined by the combination of three SO(3) vectors, such as momentum, polarization, and spin. This parameter has been widely utilized in meson and baryon decays to explore new physics-sensitive observables. The CP asymmetry induced by triple products demonstrates a unique cosine type dependence on strong phases, which has been established through a general definition and proof \cite{Wang:2022fih}. In order to determine these observables in experiments, it is essential to have more than three independent momentum variables, as polarization and spin are typically not directly measured in modern colliders. As previously mentioned, it is feasible to construct these quantities in the context of the $\Lambda\phi$ decay channels. In Table \ref{TPAs}, we have specifically presented the triple products and their corresponding asymmetries calculated in our work and in the PQCD approach. It is evident that the triple products $A^{1}_{T}$ and $\bar{A}^{1}_{T}$ exhibit significant values, primarily due to the notable strong phases associated with the helicity amplitudes. On the other hand, the remaining triple products are suppressed by the parameter $P_{b}$ that we have employed in our analysis. Overall, the triple product asymmetries are observed to be very small. This is attributed to the substantial suppression of the interference terms arising from the tree and penguin contributions as discussed in the $\Lambda\phi$ direct CP asymmetry. We therefore can conclude that the detection of significant triple product asymmetries in the $\Lambda\phi$ decay process would serve as a compelling signal of potential new physics beyond the Standard Model \cite{Bensalem:2002ys,Bensalem:2000hq,Bensalem:2002pz}.

    \item Recently, the CP asymmetries for three body decay $\Lambda^{0}_{b}\to \Lambda h^{+}h^{\prime-}$ were measured\cite{LHCb:2014yin}. This measurement shows the total CP asymmetry $\Delta \mathcal{A}_{CP}(\Lambda_b^0 \to \Lambda K^+ K^-)= 0.083\pm 0.023 \pm 0.016$ with the significance of $3.1\sigma$. More interestingly, the different resonance-dominated regions and the associated $\Delta A_{CP}$ were measured. For the $N^{*}$ dominated region, one could investigate it by applying $N\pi$ scattering mechanism as depicted in \cite{Wang:2024oyi}, while in the $\phi$ and scalar $f_{0}$ dominated regions, one can study it through hadronic re-scattering method developed in this work.  Therefore, more theoretical analysis are required in the future.
\end{itemize}

\section{Summary}\label{Summary}

We study five charmless weak decay channels of the $\Lambda^{0}_{b}$ baryon, which are expected to have potential for $CP$ violation observation. Our calculation is performed in the approach of final state interactions. Unlike its conventional approach, under which only the imaginary parts of amplitudes are taken into account by the optical theorem, our methodology involves a complete calculation of long-distance amplitudes in the form of loop integrations of triangle diagrams. It brings in the strong phases naturally and makes it possible to calculate the $CP$ asymmetries. 


We obtain the expressions of all helicity amplitudes for each decay process, incorporating various CKM components to implement the comparison between tree and penguin contributions. Our results are consistent with chiral analysis of effective weak operators and power rules of the SCET. 
A global analysis is performed to determine the two model parameters $ \Lambda_{\textbf{charm}}$ and $\Lambda_{\textbf{charmless}}$ with the experimental data on $\Lambda^{0}_{b} \to p\pi^{-}$ and $ pK^{-}$. 
Our numerical results show that the direct $CP$ asymmetries of $\Lambda_{b}^0\to p\pi^{-}$ and $pK^{-}$ decays are small because of the cancellation between  two contributing helicity amplitudes. The $CP$ asymmetry in $\Lambda_b^0 \to p\rho^-$ decay is large and may be tested in future experiments. Besides, we make predictions for branching ratios, direct $CP$ asymmetries, decaying asymmetry parameters and their associated $CP$ asymmetries, partial wave amplitude $CP$ asymmetries for $\Lambda_b^0\to p\pi^{-},pK^{-},pK^{*-},p\rho^{-},\Lambda\phi$ decays, as well as triple product correlations for  $\Lambda_b^0\to \Lambda\phi$ decay.
The branching ratio of $\Lambda_b^0 \to pK^{*-}$ is consistent with the constraint from the three-body decay $\Lambda_b^0 \to p\bar{K}^{0}\pi^{-}$\cite{LHCb:2014yin}, and the branching ratio of $\Lambda_b^0 \to \Lambda \phi$ aligns with the experimental measurement\cite{LHCb:2016hwg}. The predictions for the other observables are expected to be tested in future experiments.

Under the heavy quark symmetry and flavor $SU(3)$ symmetry, the parameters determined in this work can be borrowed by other $b$-baryon charmless decays. The formalism developed in this work can also be applied to other charmless decay channels of $\Lambda^{0}_{b}$, as well as the decays of $\Xi_{b}$ and $\Omega_{b}$. It also offers a possibility to exploring subprocesses of multibody decays such as $\Lambda^{0}_{b} \to \Lambda(1520)\phi$, $\Lambda(1520)\rho$, $N^{*}(1520)K^{*}$... and the $CP$ violating effects in multibody decays. Finally, it should be emphasized that it is difficult to take into account the effects
of $SU(3)$ flavor symmetry breaking in our calculation, as it is the basis of the strong effective Lagrangian we used. 
Currently, we adopt this approximate flavor symmetry, expecting it to provide some valuable phenomenological points.
However,it is full of challenge to fulfill the requirement of a high precision test by employing the re-scattering approach we developed.

\deleted{
It opens up avenues for exploring subprocesses of multibody decays such as $\Lambda^{0}_{b} \to \Lambda(1520)\phi, \Lambda(1520)\rho$, $N^{*}(1520)K^{*}...$, which are challenging for the perturbative QCD scheme based on factorization due to limited inputs. The formulas established in this study allow for the investigation of additional CP violating effects in the multibody decays, such as the interference effects from different intermediate resonances.} 

\section*{Acknowledgment }

The authors would like to express their gratitude to Cai-Ping Jia for helpful discussions and support with the computing program. Special thanks are extended to Wen-Jie Song for pointing out some typos in our draft. This work is supported in part by the National Key Research and Development Program of China (2023YFA1606000) and Natural Science Foundation of China under grant No. 12335003,12075126,12275277 and 12435004, by the Scientific Research Innovation Capability Support Project for Young Faculty under Grant No. ZYGXQNJSKYCXNLZCXM-P2, and by the Fundamental Research Funds for the Central Universities under No. lzujbky-2023-stlt01, lzujbky-2023-it12, lzujbky-2024-oy02 and lzujbky-2025-eyt01.
\appendix
\section{Effective Lagrangian}\label{app.A}
Effective Lagrangians for the hadronic interactions are\cite{Cheng:2004ru,Yu:2017zst,Aliev:2009ei,Aliev:2006xr}:

\begin{itemize}
    \item The effective Lagrangians for vector and pseudoscalar meson octet $V, P_8$, and baryon octet, decuplet $B_8,B_{10}$:
\begin{equation}
 \begin{aligned}
    \mathcal{L}_{V P_8 P_8}
    &=\frac{\mathrm{i} g_{\rho\pi\pi}}{\sqrt{2}} \operatorname{Tr}\left[V^\mu\left[P_8, \partial_\mu P_8\right]\right]
    \nonumber\\[3mm]
    \mathcal{L}_{V V V}
    &=\frac{i g_{\rho\rho\rho}}{\sqrt{2}} \operatorname{Tr}\left[\left(\partial_\nu V_\mu V^\mu-V^\mu \partial_\nu V_\mu\right) V^\nu\right]=\frac{\mathrm{i} g_{\rho\rho\rho}}{\sqrt{2}} \operatorname{Tr}\left[\left(\partial_\nu V_\mu-\partial_\mu V_\nu\right) V^\mu V^\nu\right]
    \nonumber\\[3mm]
    \mathcal{L}_{V V P_8}
    &=\frac{4 g_{V V P_8}}{f_{P_8}} \varepsilon^{\mu \nu \alpha \beta} \operatorname{Tr}\left(\partial_\mu V_\nu \partial_\alpha V_\beta P_8\right)
    \nonumber\\[3mm]
    \mathcal{L}_{P_8 B_8 B_8}
    &=\sqrt{2}\Big(D\operatorname{Tr}\left[ \bar{B_8} \{P_8,B_8\} \right]+F\operatorname{Tr}\left[ \bar{B_8} [P_8,B_8]\right] \Big)\nonumber\\[3mm]
    \mathcal{L}_{V B_8 B_8}
    &=\sqrt{2}\Big(F \operatorname{Tr} \left[ \bar{B} [V,B_8] \right]+D \operatorname{Tr}\left[ \bar{B_8} \{V,B_8\} \right]+(F-D)\operatorname{Tr}[\bar{B_8} B_8]\;\operatorname{Tr}[V]\Big)\nonumber\\[3mm]
    \mathcal{L}_{P_8 B_8 B_{10}}
    &=\frac{g_{\pi N\Delta}}{m_{\pi}}\bar{B}^{\mu}_{10} \partial_\mu P_8 B_8+h.c.
    \nonumber\\[3mm]
    \mathcal{L}_{V B_8 B_{10}}
    &=-i \frac{g_{\rho N\Delta}}{m_\rho}\bar{B}^{\mu}_{10} \gamma^5 \gamma^\nu B_8\left(\partial_\mu V_\nu -\partial_\nu V_\mu\right)+h.c. \nonumber
 \end{aligned}
 \end{equation}
 
\item The Lagrangians involving $D^{(*)}$-mesons, pseudoscalar meson octet $P_8$ and vector meson octet $V$ :
\begin{equation}
 \begin{aligned}
    \mathcal{L}_{D^\ast DP_8}
    &=-i g_{D^* D P_8}\left(D^i \partial^\mu P_{8 i j} D_\mu^{* j \dagger}-D_\mu^{* i} \partial^\mu P_{8 i j} D^{j \dagger}\right)\nonumber\\[3mm]
    \mathcal{L}_{D^\ast D^\ast P_8}
    &=\frac{1}{2} g_{D^* D^* P_8} \varepsilon_{\mu \nu \alpha \beta} D_i^{* \mu} \partial^\nu P^{i j}_8 \overleftrightarrow{\partial}^\alpha D_j^{* \beta \dagger}\nonumber\\[3mm]
    \mathcal{L}_{DDV}
    &=-i g_{D D V} D_i\overleftrightarrow{\partial}_\mu D^{j\dagger}\left(V^\mu\right)_j^i\nonumber\\[3mm]
    \mathcal{L}_{D^\ast DV}
    &=-2 f_{D^* D V} \varepsilon_{\mu \nu \alpha \beta}\left(\partial^\mu V^\nu\right)_j^i\left(D_i \overleftrightarrow{\partial}^\alpha D^{* \beta j\dagger}-D_i^{* \beta } \overleftrightarrow{\partial}^\alpha D^{j\dagger}\right)\nonumber\\[3mm]
    \mathcal{L}_{D^\ast D^\ast V}&=i g_{D^* D^* V} D_i^{* \nu } \overleftrightarrow{\partial}_\mu D_\nu^{* j\dagger}\left(V^\mu\right)_j^i+4 i f_{D^* D^* V} D_{i \mu}^{* }\left(\partial^\mu V^\nu-\partial^\nu V^\mu\right)_j^i D_\nu^{* j\dagger}\nonumber
 \end{aligned}
 \end{equation}

 \item The Lagrangians involving charmed baryon sextets $B_6$, anti-triplets $B_{\bar{3}}$, vector and pseudo-scalar mesons octet $V,P_8$:
 \begin{equation}
 \begin{aligned}
    \mathcal{L}_{Vhh}
    &=\left\{f_{1V {B}_6 {B}_6} \operatorname{Tr}\left[\bar{{B}}_6 \gamma_\mu V^\mu {B}_6\right]+\frac{f_{2 V {B}_6 {B}_6}}{m_6+m^{\prime}_{6} }\operatorname{Tr}\left[\bar{{B}}_6 \sigma_{\mu \nu} \partial^\mu V^\nu {B}_6\right] \right\}\nonumber\\
    & +\left\{f_{1 V {B}_{\bar{3}} {B}_{\bar{3}}} \operatorname{Tr}\left[\bar{{B}}_{\bar{3}} \gamma_\mu V^\mu {B}_{\bar{3}}\right]+\frac{f_{2 V {B}_{\bar{3}} {B}_{\bar{3}}}}{ m_{\bar{3}}+m_{\bar{3}}^\prime} \operatorname{Tr}\left[\bar{{B}}_{\bar{3}} \sigma_{\mu \nu} \partial^\mu V^\nu {B}_{\bar{3}}\right]\right\}\nonumber \\
    & +\left\{f_{1 V {B}_6 {B}_{\bar{3}}} \operatorname{Tr}\left[\bar{{B}}_6 \gamma_\mu V^\mu {B}_{\bar{3}}\right]+\frac{f_{2 V {B}_6 {B}_{\bar{3}}}}{m_6+m_{\bar{3}}} \operatorname{Tr}\left[\bar{{B}}_6 \sigma_{\mu \nu} \partial^\mu V^\nu {B}_{\bar{3}}\right]+h . c .\right\}\nonumber \\[3mm]
     \mathcal{L}_{Phh}
     &=g_{P_8 {B}_6 {B}_6} \operatorname{Tr}\left[\bar{{B}}_6 i \gamma_5 P_8 {B}_6\right]+g_{P_8 {B}_{\bar{3}} {B}_{\bar{3}}} \operatorname{Tr}\left[\bar{{B}}_{\bar{3}} i \gamma_5 P_8 {B}_{\bar{3}}\right]+\left\{g_{P_8 {B}_6 {B}_{\bar{3}}} \operatorname{Tr}\left[\bar{{B}}_6 i \gamma_5 P_8 {B}_{\bar{3}}\right]+h . c .\right\}\nonumber\nonumber
 \end{aligned}
 \end{equation}
 \item The Lagrangians for charmed baryon sextets  $B_6$, anti-triplets $B_{\bar{3}}$, baryon octet $B_8$ and $D^{(*)}$-mesons:
  \begin{equation}
 \begin{aligned}
     \mathcal{L}_{\Lambda_{\mathrm{c}} N D}
     &=g_{\Lambda_{\mathrm{c}} N D}\left(\bar{\Lambda}_{\mathrm{c}} i \gamma_5 D N+{ h.c. }\right)\nonumber\\[3mm]
     \mathcal{L}_{\Lambda_{\mathrm{c}} N D^*}
     &=f_{1 \Lambda_{\mathrm{c}} N D^*}\left(\bar{\Lambda}_{\mathrm{c}} \gamma_\mu D^{* \mu} N+ { h.c. }\right)+\frac{f_{2 \Lambda_{\mathrm{c}} N D^*}}{m_{\Lambda_{\mathrm{c}}}+m_N}\left(\bar{\Lambda}_{\mathrm{c}} \sigma_{\mu \nu} \partial^\mu D^{* \nu} N+{ h.c. }\right),\nonumber\\[3mm]
     \mathcal{L}_{\Sigma_{\mathrm{c}} N D}
     &= g_{\Sigma_{\mathrm{c}} N D}\left(\bar{\Sigma}_{\mathrm{c}} i \gamma_5 D N+{ h.c. }\right)\nonumber\\[3mm]
     \mathcal{L}_{\Sigma_{\mathrm{c}} N D^*}
     & =f_{1 \Sigma_{\mathrm{c}} N D^*}\left(\bar{\Sigma}_{\mathrm{c}} \gamma_\mu D^{* \mu} N+{ h.c. }\right)+\frac{f_{2 \Sigma_{\mathrm{c}} N D^*}}{m_{\Sigma_{\mathrm{c}}}+m_N}\left(\bar{\Sigma}_{\mathrm{c}} \sigma_{\mu \nu} \partial^\mu D^{* \nu} N+{ h.c. }\right)\nonumber
\end{aligned}
 \end{equation}
 \end{itemize}

The matrices under SU(3) flavor group representations are given:

  \begin{equation}
 \begin{aligned}
P=\left(\begin{array}{ccc}
\frac{\pi^0}{\sqrt{2}}+\frac{\eta}{\sqrt{6}} & \pi^{+} & \mathrm{K}^{+} \\
\pi^{-} & -\frac{\pi^0}{\sqrt{2}}+\frac{\eta}{\sqrt{6}} & \mathrm{~K}^0 \\
\mathrm{~K}^{-} & \bar{K}^0 & -\sqrt{\frac{2}{3}} \eta
\end{array}\right), \quad {B}_6=\left(\begin{array}{ccc}
\Sigma_{\mathrm{c}}^{++} & \frac{1}{\sqrt{2}} \Sigma_{\mathrm{c}}^{+} & \frac{1}{\sqrt{2}} \Xi_{\mathrm{c}}^{\prime+} \\
\frac{1}{\sqrt{2}} \Sigma_{\mathrm{c}}^{+} & \Sigma_{\mathrm{c}}^0 & \frac{1}{\sqrt{2}} \Xi_{\mathrm{c}}^{\prime 0} \\
\frac{1}{\sqrt{2}} \Xi_{\mathrm{c}}^{\prime+} & \frac{1}{\sqrt{2}} \Xi_{\mathrm{c}}^{\prime 0} & \Omega_{\mathrm{c}}
\end{array}\right),
\end{aligned}
 \end{equation}

   \begin{equation}
 \begin{aligned}
V=\left(\begin{array}{ccc}
\frac{\rho^0}{\sqrt{2}}+\frac{\omega}{\sqrt{2}} & \rho^{+} & \mathrm{K}^{*+} \\
\rho^{-} & -\frac{\rho^0}{\sqrt{2}}+\frac{\omega}{\sqrt{2}} & \mathrm{~K}^{* 0} \\
\mathrm{~K}^{*-} & \bar{\mathrm{K}}^{* 0} & \phi
\end{array}\right), ~~~~~~~~~~~~~\quad {B}_{\bar{3}}=\left(\begin{array}{ccc}
0 & \Lambda_{\mathrm{c}}^{+} & \Xi_{\mathrm{c}}^{+} \\
-\Lambda_{\mathrm{c}}^{+} & 0 & \Xi_{\mathrm{c}}^0 \\
-\Xi_{\mathrm{c}}^{+} & -\Xi_{\mathrm{c}}^0 & 0
\end{array}\right),
\end{aligned}
 \end{equation}
 
  \begin{equation}
 \begin{aligned}
 {B_8}=\left(\begin{array}{ccc}
\frac{\Sigma^0}{\sqrt{2}}+\frac{\Lambda}{\sqrt{6}} & \Sigma^{+} & p \\
\Sigma^{-} & -\frac{\Sigma^0}{\sqrt{2}}+\frac{\Lambda}{\sqrt{6}} & n \\
\Xi^{-} & \Xi^0 & -\frac{2}{\sqrt{6}} \Lambda
\end{array}\right)
 ,~~~~~~~~~~~~~~~\quad D=\left(
\begin{matrix}
    D^0,D^+,D_s^+
\end{matrix}\right)
\end{aligned}
 \end{equation}

\section{The Feynman rules of strong vertex }\label{app.Bprime}
\begin{equation}
	\begin{aligned}
\bra{P_8(p_{3})D(k,\lambda_{k})}i\mathcal{L}\ket{D^{\ast}(p_{1},\lambda_{1})}
    &=ig_{D^{\ast} D P_8}p^{\mu}_{3}\varepsilon_{\mu}(p_1,\lambda_1)\\
\bra{P_8(p_{3})D^*(k,\lambda_{k})}i\mathcal{L}\ket{D^{\ast}(p_1,\lambda_{1})}
    &= \frac{i}{2}g_{D^*D^*P_8}\varepsilon_{\mu \nu \alpha \beta}\varepsilon^{*\mu}(k,\lambda_k)\varepsilon^\beta(p_1,\lambda_1)p^\nu_{3} p^\alpha_{1}\\
\bra{V(p_{3},\lambda_{3})D^*(k,\lambda_{k})}i\mathcal{L}\ket{D^*(p_{1},\lambda_{1})}
    &=2ig_{D^*D^*V}\varepsilon^{*\nu}(k,\lambda_k)\varepsilon_{\nu}(p_1,\lambda_1)k_\mu \varepsilon^{*\mu}(p_{3},\lambda_3)\\
    &-4if_{D^* D^*V}\varepsilon_{\mu}^*(k,\lambda_k) \Big(p^\mu_{3} \varepsilon^{*\nu}(p_{3},\lambda_3)-p^\nu_{3}\varepsilon^{*\mu}(p_{3},\lambda_3)\Big)\varepsilon_{\nu}(p_{1},\lambda_1)\\
\bra{V(p_{3},\lambda_{3})D^*(k,\lambda_{k})}i\mathcal{L}\ket{D(p_{1})}
   & =2 i f_{D^*DV} \varepsilon_{\mu \nu \alpha \beta} \varepsilon^{*\nu}(p_{3},\lambda_3) \varepsilon^{*\beta}(k,\lambda_k) p^\mu_{3} (k^\alpha +p^\alpha_{1})\\
\bra{V(p_{3},\lambda_{3})D(k)}i\mathcal{L}\ket{D(p_{1})}
    &=-i g_{DDV}\varepsilon^{*\mu}(p_{3},\lambda_3)(p_{1,\mu}+k_{\mu}),
    \end{aligned}
\end{equation}
\begin{equation}
	\begin{aligned}\label{eq:strong vertex1}
\langle V(k,\lambda_k)P(p_{2})|i\mathcal{L}_{VPP}|P(p_{1})\rangle&=-ig_{VVP}\varepsilon^{*\mu}(k,\lambda_k)(p_{1}+p_{2})_{\mu}\,,\\
\langle\mathcal{B}_{2}(p_{2})P(q)|i\mathcal{L}_{PBB}|\mathcal{B}_{1}(p_{1})\rangle&=g_{BBP}\bar{u}(p_{2})i\gamma_{5}u(p_{1})\,,\\
\langle\mathcal{B}_{2}(p_{2})V(q,\lambda_q)|i\mathcal{L}_{VBB}|\mathcal{B}_{1}(p_{1})\rangle&=\bar{u}(p_{2})\left[f_{1}\gamma_{\nu}+f_{2}\frac{i}{m_{1}+m_{2}}\sigma_{\mu\nu}q^{\mu}\right]\varepsilon^{*\nu}(q,\lambda_q)u(p_{1})\,,  \\
\langle V(p_{3},\lambda_{3})V(k,\lambda_{k})|i\mathcal{L}_{VVP}|P(p_{1})\rangle&=-i\frac{g_{VVP}}{f_{p}}\epsilon^{\mu\nu\alpha\beta}p_{3\,\mu}\varepsilon^{*}_{\nu}(\lambda_{3},p_{3})k_{\alpha}\varepsilon^{*}_{\beta}(k,\lambda_{k})\,,\\
\langle B(p_4,\lambda_{4})|i\mathcal{L}_{VBD}|D(k,\lambda_{k})V(p_{1},\lambda_{1})\rangle &=-i\frac{g_{\rho N\Delta}}{m_{\rho}}\bar{u}(p_{4},\lambda_{4})\gamma^{5}\gamma^{\nu}u^{\mu}(k,\lambda_{k})\\
&\times\left[p_{1\,\mu}\varepsilon_{\nu}(p_1,\lambda_1)-p_{1\,\nu}\varepsilon_{\mu}(p_1,\lambda_1)\right]\,, \\
\langle B(p_{4},\lambda_{4})|i\mathcal{L}_{VBD}|D(k,\lambda_{k})P(p_{1},\lambda_{1})\rangle
&=\frac{g_{\pi N\Delta}}{m_{\pi}} \bar{u}(p_{4},\lambda_{4})p_{1\,\mu}u^{\mu}(k,\lambda_{k})\,,
    \end{aligned}
\end{equation}
\begin{equation}
	\begin{aligned}
\label{eq:strong vertex2}
\langle V(p_{3},\lambda_{3})V(k,\lambda_{k})|i\mathcal{L}_{VVV}|V(p_{1},\lambda_{1})\rangle
&= -\frac{ig_{VVV}}{\sqrt{2}}\varepsilon_{\mu}(p_{1},\lambda_{1})\varepsilon^{\mu*}(p_{3},\lambda_{3})\varepsilon^{*}_{\nu}(k,\lambda_{k})\left(p^{\nu}_{3}+p^{\nu}_{1}\right)\\
        & -\frac{ig_{VVV}}{\sqrt{2}}\varepsilon^{*}_{\mu}(k,\lambda_{k})\varepsilon^{\mu}(p_{1},\lambda_{1})\varepsilon^{*}_{\nu}(p_{3},\lambda_{3})\left(-p^{\nu}_{1}-p^{\nu}_{k}\right)\\
        &-\frac{ig_{VVV}}{\sqrt{2}}\varepsilon^{*}_{\mu}(p_{3},\lambda_{3})\varepsilon^{*\mu}(k,\lambda_{k})\varepsilon_{\nu}(p_{1},\lambda_{1})\left(p^{\nu}_{k}-p^{\nu}_{3}\right).
    \end{aligned}
\end{equation}

\section{Amplitudes of triangle diagram}\label{app.B}
The amplitudes of $\Lambda^{0}_{b} \rightarrow B_{8} P_{8} $:
\begin{equation}
	\begin{aligned}
 \mathcal{M}[P_8,B_8;V]&= \int\frac{d^{4}k}{(2\pi)^{4}}(-1) g_{V P_8 P_8}\cdot\bar{u} 
       (p_4,\lambda_4)(f_{1V B_8 B_8}\cdot\gamma_{\mu}-\frac{if_{2V B_8B_8}}{m_2+m_4}\sigma_{\nu\mu}k^{\nu})(\not\! p_2+m_2)(A+B\gamma_5)u(p_i,\lambda_i)\\
       &(-g^{\alpha \mu}+\frac{k^{\alpha}k^\mu}{m_{k}^2})(p_{1\alpha}+p_{3\alpha})\frac{\mathcal{F}}{(p_1^2-m_1^2+i\varepsilon)(p_2^2-m_2^2+i\varepsilon)(k^2-m_k^2+i\varepsilon)}
       	\end{aligned}
\end{equation}

\begin{equation}
	\begin{aligned}
\mathcal{M}[P_8,B_8;B_8]
       &=\int \frac{d^4k}{(2\pi)^4}g_{P_8 B_8 B_8}\cdot g_{P_8 B_8 B_8}^\prime \bar{u}(p_4,\lambda_4)\gamma_5(\not\!k+m_k)\gamma_5(\not\!p_2+m_2)(A+B\gamma_5)u(p_i,\lambda_i)\\
      & \cdot\frac{1}{(p_1^2-m_1^2+i\varepsilon)}
       \cdot\frac{\mathcal{F}}{(p_2^2-m_2^2+i\varepsilon)(k^2-m_k^2+i\varepsilon)}
       	\end{aligned}
\end{equation}

\begin{equation}
 \begin{aligned}
\mathcal{M}[V,B_8;P_8]
       &=\int \frac{d^4k}{(2\pi)^4} g_{P_8 B_8 B_8}\cdot g_{VP_8P_8} \bar{u}(p_4,\lambda_4)\gamma_5(\not\!p_2+m_2)(A_1\gamma_{\mu}\gamma_5+A_2\frac{p_{2\mu}}{m_i}\gamma_5+B_1\gamma_{\mu}+B_2\frac{p_{2\mu}}{m_i})u(p_i,\lambda_i)\\
       &(-g^{\mu\nu}+\frac{p_1^{\mu} p_1^{\nu}}{m_1^2})(p_{3\nu}-k_{\nu})\frac{\mathcal{F}}{(p_1^2-m_1^2+i\varepsilon)(p_2^2-m_2^2+i\varepsilon)(k^2-m_k^2+i\varepsilon)}
	\end{aligned}
\end{equation}

\begin{equation}
 \begin{aligned}
\mathcal{M}[V,B_8;V]
       &=\int \frac{d^4k}{(2\pi)^4}i\cdot \frac{4g_{V V P_8}}{f_{P_8}}\cdot\bar{u}(p_4,\lambda_4)\left(f_{1V BB}\cdot\gamma^{\sigma}-\frac{if_{2V BB}}{m_2+m_4}\cdot\sigma^{\nu\sigma}k_{\nu}\right)(\not\!p_2 +m_2)\cdot\bigg(A_1\gamma^{\delta}\gamma_5\\
       &+A_2\frac{p_2^{\delta}} {m_i}\gamma_5
       +B_1\gamma^{\delta} +B_2\frac{p_2^{\delta}}{m_i}\bigg)u(p_i,\lambda_i)p_1^{\beta}k^{\rho}\varepsilon_{\beta\delta\rho\sigma}\cdot\frac{\mathcal{F}}{(p_1^2-m_1^2+i\varepsilon)(p_2^2-m_2^2+i\varepsilon)(k^2-m_k^2+i\varepsilon)}
	\end{aligned}
\end{equation}

\begin{equation}
 \begin{aligned}
\mathcal{M}[V,B_8;B_8]
       &=\int \frac{d^4k}{(2\pi)^4}(-1)g_{P_8 B_8 B_8}\cdot\bar{u} (p_4,\lambda_4)
       \left(f_{1V BB}\cdot\gamma_{\mu}-\frac{if_{2V BB}} 
       {m_k+m_4}\cdot\sigma_{\nu\mu}p_{1}^{\nu}\right)\left(\not\!k +m_k\right)\gamma_5\left(\not\!p_2+m_2\right)
       \bigg(A_1\gamma_{\alpha}\gamma_5\\
       &+A_2\frac{p_{2\alpha}}{m_i}\gamma_5+B_1\gamma_{\alpha}+B_2\frac{p_{2\alpha}}{m_i}\bigg)u(p_i,\lambda_i) 
       \left(-g^{\alpha\mu}+\frac{p_1^{\alpha}p_1^{\mu}}{m_1^2}\right)\frac{\mathcal{F}}{(p_1^2-m_1^2+i\varepsilon)(p_2^2-m_2^2+i\varepsilon)(k^2-m_k^2+i\varepsilon)}
	\end{aligned}
\end{equation}

\begin{equation}
 \begin{aligned}
\mathcal{M}[P_8,B_8;B_{10}]
       &=\int \frac{d^4k}{(2 \pi)^4}(-\frac{1}{m_1 m_3}\cdot g_{1 P_8 B_8 B_{10}}\cdot g_{2 P_8 B_8 B_{10}})\cdot\bar{u}(p_4,\lambda_4)(\not\!k+m_k)\cdot\Big\{-g_{\mu\nu}+\frac{1}{3}\gamma_{\mu}\gamma_{\nu}+\frac{2}{3m_{k}^2}k_{\mu}k_{\nu}\\
       &-\frac{1}{3m_k}(k_{\mu}\gamma_{\nu}-k_{\nu}\gamma_{\mu})\Big\}(\not\!p_2+m_2)(A+B\gamma_5)u(p_i,\lambda_i)p_1^{\mu}p_3^{\nu}\cdot\frac{\mathcal{F}}{(p_1^2-m_1^2+i\varepsilon)(p_2^2-m_2^2+i\varepsilon)(k^2-m_k^2+i\varepsilon)}
	\end{aligned}
\end{equation}

\begin{equation}
 \begin{aligned}
\mathcal{M}[V,B_8;B_{10}]
        &=\int \frac{d^4k}{(2\pi)^4}\frac{g_{PB_8 B_{10}}\cdot g_{VB_8 B_{10}}}{m_1\cdot m_3}\cdot\bar{u}(p_4,\lambda_4)\gamma_5\gamma_{\alpha}(\not\!k +m_k) \Big\{-g_{\mu\nu}+\frac{1}{3}\gamma_{\mu}\gamma_{\nu}+\frac{2}{3m_k^2}k_{\mu}k_{\nu}-\frac{1}{3m_k}\left(k_{\mu}\gamma_{\nu}-k_{\nu}\gamma_{\mu}\right)\Big\}\\
        &(\not\!p_2 +m_2)\left(A_1\gamma_{\beta}\gamma_5+A_2\frac{p_{2\beta}}{m_i}\gamma_5 +B_1\gamma_{\beta}+B_2\frac{ p_{2\beta}}{m_i}\right)u(p_i,\lambda_i)\left(p_1^{\alpha}g^{\mu\beta}-p_1^{\mu}g^{\alpha\beta}\right)p_3^{\nu}\cdot\frac{1}{(p_1^2-m_1^2+i\varepsilon)}\\
        &\cdot\frac{\mathcal{F}}{(p_2^2-m_2^2+i\varepsilon)(k^2-m_k^2+i\varepsilon)}
    \end{aligned}
\end{equation}

\begin{equation}
 \begin{aligned}
\mathcal{M}[D^{\ast},B_{\bar{3}},D]
        &=\int \frac{d^4k}{(2\pi)^4}(-1)g_{B_{\bar{3}} B_8 D}\cdot g_{ D^{\ast}D P_8}\bar{u}(p_4,\lambda_4)\gamma_5(\not\!p_2+m_2)\left(A_1\gamma_{\mu}\gamma_5+A_2\frac{p_{2\mu}}{m_{i}}\gamma_5+B_1\gamma_{\mu}+B_2\frac{p_{2\mu}}{m_{i}}\right)u(p_i,\lambda_i)\\
        &(-g^{\mu\nu}+\frac{p_1^{\mu}p_1^{\nu}}{m_1^2})p_{3\nu}\cdot\frac{\mathcal{F}}{(p_1^2-m_1^2+i\varepsilon)(p_2^2-m_2^2+i\varepsilon)(k^2-m_k^2+i\varepsilon)}
    \end{aligned}
\end{equation}

\begin{equation}
 \begin{aligned}
\mathcal{M}[D,B_{\bar{3}},D^{\ast}]
        &=\int \frac{d^4k}{(2\pi)^4}(-1)g_{D^{\ast} DP_8}\cdot \bar{u}(p_4,\lambda_4)\left(f_{1B_{\bar{3}} B_8 D^{\ast}}\cdot\gamma_{\mu}-\frac{if_{2B_{\bar{3}} B D^{\ast}}}{m_2+m_4}\cdot \sigma_{\alpha\mu}k^{\alpha}\right)(\not\!p_2+m_2)(A+B\gamma_5)u(p_i,\lambda_i)\\
        &(-g^{\mu\nu}+\frac{k^{\mu}k^{\nu}}{m_k^2})p_{3\nu}\cdot\frac{\mathcal{F}}{(p_1^2-m_1^2+i\varepsilon)(p_2^2-m_2^2+i\varepsilon)(k^2-m_k^2+i\varepsilon)}
    \end{aligned}
\end{equation}

The general formalism of $\Lambda^{0}_{b} \rightarrow B_8 V$ amplitudes:
\begin{equation}
 \begin{aligned}
\mathcal{M}[P_8,B_8;P_8]
       &=\int\frac{d^4k}{(2 \pi)^4}ig_{P_8B_8B_8}\cdot g_{VP_8 P_8}\bar{u}(p_4,\lambda_4)\gamma_5 (\not\!p_2+m_2)(A+B\gamma_5)u(p_i,\lambda_i)\varepsilon^{*\mu}(p_3,\lambda_3)(k_\mu+p_{1\mu})\\
       &\cdot\frac{\mathcal{F}}{(p_1^2-m_1^2+i\varepsilon)(p_2^2-m_2^2+i\varepsilon)(k^2-m_k^2+i\varepsilon)}
    \end{aligned}
\end{equation}

\begin{equation}
 \begin{aligned}
\mathcal{M}[V,B_8;P_8]
     &=\int \frac{d^4k}{(2\pi)^4}(-1)g_{P_8 B_8 B_8 }\cdot \frac{4g_{P_8VV}}{f_p}\varepsilon^{\mu \nu \alpha \beta}\varepsilon^*_{\beta}(p_3,\lambda_3) 
     p_{1\mu} p_{3\alpha}(-g_{\nu\delta}+\frac{p_{1\nu} p_{1\delta}}{m_1^2})\bar{u}(p_4,\lambda_4)\gamma_5(\not\!p_2+m_2)\\
     &\cdot (A_1\gamma^\delta \gamma_5+A_2 \frac{p_2^\delta}{m_i}\gamma_5+B_1\gamma_\delta+B_2\frac{p_2^\delta}{m_i})u(p_i,\lambda_i)\cdot\frac{\mathcal{F}}{(p_1^2-m_1^2+i\varepsilon)(p_2^2-m_2^2+i\varepsilon)(k^2-m_k^2+i\varepsilon)}
\end{aligned}
\end{equation}

\begin{equation}
 \begin{aligned}
\mathcal{M}[P_8,B_8;V]
     &=\int\frac{d^4k}{(2\pi)^4}\frac{4g_{P_8VV}}{f_{P_8}}\bar{u}(p_4,\lambda_4)(f_{1VB_8B_8}\gamma_\delta-\frac{if_{2VB_8B_8}}{m_2+m_4}\sigma_{\rho\delta}k^\rho)(\not\!p_2+m_2)(A+B\gamma_5)u(p_i,\lambda_i)\\
     &(-g^{\delta\nu}+\frac{k^\delta k^\nu}{m_k^2})\varepsilon^{\mu \nu \alpha \beta}\varepsilon_{\beta}^*(p_3,\lambda_3)k_{\mu}p_{3\alpha}\cdot\frac{\mathcal{F}}{(p_1^2-m_1^2+i\varepsilon)(p_2^2-m_2^2+i\varepsilon)(k^2-m_k^2+i\varepsilon)}
         \end{aligned}
\end{equation}

     \begin{equation}
 \begin{aligned}
\mathcal{M}[V,B_8,V]
     &=\int\frac{d^4k}{(2\pi)^4}(-\frac{i}{\sqrt{2}})g_{VVV}\bar{u}(p_4,\lambda_4)(f_{1VB_8B_8}\gamma_\alpha-\frac{if_{2VB_8B_8}}{m_2+m_4}\sigma_{\beta\alpha}k^\beta)(\not\!p_2+m_2)\cdot \bigg(A_1\gamma^\delta\gamma_5+A_2\frac{p_2^\delta}{m_i}\gamma_5+B_1\gamma^\delta\\
     &+B_2\frac{p_2^\delta}{m_i}\bigg)u(p_i,\lambda_i)\cdot \bigg\{2k_\nu\varepsilon^{*\nu}(p_3,\lambda_3)(-g_{\mu\delta}+\frac{p_{1\mu}p_{1\delta}}{m_1^2})(-g^{\alpha\mu}+\frac{k^\alpha k^\mu }{m_k^2})+\Big(-p_{1\nu}\varepsilon^{*\mu}(p_3,\lambda_3)+p^{\mu}_3\varepsilon_{\nu}^*(p_3,\lambda_3)\\ 
     &-p_{3\nu}\varepsilon^{*\mu}(p_3,\lambda_3)-k^\mu \varepsilon^*_{\nu}(p_3,\lambda_3) \Big) \cdot(-g_{\mu\delta}+\frac{p_{1\mu}p_{1\delta}}{m_1^2})(-g^{\alpha \nu}+\frac{k^\alpha k^\nu}{m_k^2})  \bigg\} \cdot\frac{1}{(p_1^2-m_1^2+i\varepsilon)(p_2^2-m_2^2+i\varepsilon)}\\
     &\cdot \frac{\mathcal{F}}{(k^2-m_k^2+i\varepsilon)}
         \end{aligned}
\end{equation}

  \begin{equation}
 \begin{aligned}   
     \mathcal{M}[P_8,B_8;B_8]
     &=\int\frac{d^4k}{(2\pi)^4}(-i)g_{P_8 B_8 B_8}\bar{u}(p_4,\lambda_4)\gamma_5(\not\!k+m_k)\Big(f_{1VB_8B_8}\gamma_\mu+\frac{if_{2VB_8B_8}}{m_2+m_k}\sigma_{\nu\mu}p_3^\nu \Big)(\not\!p_2+m_2)(A+B\gamma_5)\\
     &\cdot \varepsilon^{*\mu}(p_3,\lambda_3) u(p_i,\lambda_i)\cdot\frac{\mathcal{F}}{(p_1^2-m_1^2+i\varepsilon)(p_2^2-m_2^2+i\varepsilon)(k^2-m_k^2+i\varepsilon)}
         \end{aligned}
\end{equation}

     \begin{equation}
 \begin{aligned}
\mathcal{M}[V,B_8;B_8]
     &=\int \frac{d^4k}{(2\pi)^4}i\bar{u}(p_4,\lambda_4)\Big(f_{1VB_8B_8}\gamma_\mu-\frac{if_{2VB_8B_8}}{m_k+m_4}\sigma_{\nu\mu}p_1^\nu\Big)(\not\!k+m_k)\Big(f^\prime_{1VB_8B_8}\gamma_\alpha+\frac{if^\prime_{2VB_8B_8}}{m_2+m_k}\sigma_{\beta\alpha}p_3^\beta \Big)\\
     &\cdot (-g^{\mu\delta}+\frac{p_1^\delta p_1^\mu}{m_1^2})\varepsilon^{*\alpha}(p_3,\lambda_3)(\not\!p_2+m_2)\Big(A_1\gamma_\delta\gamma_5+A_2\frac{p_{2\delta}}{m_i}\gamma_5+B_1\gamma_\delta+B_2\frac{p_{2\delta}}{m_i}\Big)u(p_i,\lambda_i)\\
     &\cdot\frac{1}{(p_1^2-m_1^2+i\varepsilon)}\cdot \frac{\mathcal{F}}{(p_2^2-m_2^2+i\varepsilon)(k^2-m_k^2+i\varepsilon)}
    \end{aligned}
\end{equation}
     
     \begin{equation}
 \begin{aligned}
\mathcal{M}[D^*,B_{\bar{3}};D^*]
    &=\int\frac{d^4k}{(2\pi)^4}i\bar{u}(p_4,\lambda_4)\Big(f_{1VB_8B_8}\gamma_\alpha-\frac{if_{2VB_8B_8}}{m_2+m_4}\sigma_{\beta\alpha}k^\beta\Big)(\not\!p_2+m_2)\Big(A_1\gamma_\delta \gamma_5+A_2\frac{p_{2\delta}}{m_i}\gamma_5+B_1\gamma_\delta+B_2\frac{p_{2\delta}}{m_i}\Big)\\
    &\cdot u(p_i,\lambda_i)\cdot\bigg\{2g_{D^*D^*V}(-g_{\nu\delta}+\frac{p_{1\nu }p_{1\delta}}{m_1^2})(-g^{\nu \alpha}+\frac{k^\alpha k^\nu}{m_k^2})\varepsilon^{*\mu}(p_3,\lambda_3)k_\mu-4f_{D^*D^*V}(-g_{\nu \delta}+\frac{p_{1\nu}p_{1\delta}}{m_1^2})\\
    &\cdot(-g^{\mu \alpha}+\frac{k^\mu k^\alpha}{m_k^2})\Big(p_3^\mu\varepsilon^{*\nu}(p_3,\lambda_3)-p_3^\nu \varepsilon^{*\mu}(p_3,\lambda_3)\Big)\bigg\}\cdot\frac{\mathcal{F}}{(p_1^2-m_1^2+i\varepsilon)(p_2^2-m_2^2+i\varepsilon)(k^2-m_k^2+i\varepsilon)}
        \end{aligned}
\end{equation}

         \begin{equation}
 \begin{aligned}
\mathcal{M}[P_8,B_8;B_{10}]
    &=\int \frac{d^4k}{(2\pi)^4}i\frac{g_{P_8B_8B_{10}}}{m_{1}}\frac{g_{VB_8B_{10}}}{m_3}\bar{u}(p_4,\lambda_4)p_1^\alpha(\not\!k+m_k)\bigg\{-g_{\alpha\mu}+\frac{1}{3}\gamma_\alpha \gamma_\mu +\frac{2}{3m_k^2}k_\alpha k_\mu -\frac{1}{3m_k}(k_\alpha \gamma_\mu-k_\mu \gamma_\alpha)\bigg\}\\
    &\cdot\gamma_5\gamma^\nu (\not\!p_2+m_2)\Big(p_3^{\mu}\varepsilon_{\nu}^*(p_3,\lambda_3)-p_{3\nu}\varepsilon^{* \mu }(p_3,\lambda_3)\Big)(A+B\gamma_5)u(p_i,\lambda_i)\cdot\frac{1}{(p_1^2-m_1^2+i\varepsilon)(p_2^2-m_2^2+i\varepsilon)}\\
    &\cdot \frac{\mathcal{F}}{(k^2-m_k^2+i\varepsilon)}
        \end{aligned}
\end{equation}

         \begin{equation}
 \begin{aligned}
\mathcal{M}[V,B_8;B_{10}]
    &=\int \frac{d^4k}{(2\pi)^4}(-i)\frac{g_{VB_8 B_{10}}}{m_1}\frac{g_{VB_8 B_{10}}^\prime}{m_3}\bar{u}(p_4,\lambda_4)\gamma_5\gamma_\nu(\not\!k+m_k)\cdot \bigg\{-g_{\mu\alpha}+\frac{1}{3}\gamma_\mu\gamma_\alpha+\frac{2}{3m_k^2}k_\mu k_\alpha\\
    &- \frac{1}{3m_k}\Big(k_\mu \gamma_\alpha-k_\alpha\gamma_\mu\Big)\bigg\}\cdot \gamma_5\gamma^\beta\Big(p_{3\alpha}\varepsilon_{\beta}^*(p_3,\lambda_3)-p_{3\beta}\varepsilon_{\alpha}^*(p_3,\lambda_3)\Big)(\not\!p_2+m_2)\bigg\{p_1^\mu(-g^{\delta\nu}+\frac{p_{1}^\delta p_{1}^\nu}{m_1^2})\\
    &-p_1^\nu(-g^{\mu\delta}+\frac{p_1^\delta p_1^\mu}{m_1^2})\bigg\}\cdot \bigg\{A_1\gamma_\delta \gamma_5+A_2\frac{p_{2\delta}}{m_i}\gamma_5+B_1\gamma_\delta+B_2\frac{p_{2\delta}}{m_i}\bigg\}u(p_i,\lambda_i)\cdot\frac{1}{(p_1^2-m_1^2+i\varepsilon)(p_2^2-m_2^2+i\varepsilon)}\\
    &\cdot\frac{\mathcal{F}}{(k^2-m_k^2+i\varepsilon)}
    \end{aligned}
\end{equation}
In the above complete derivation, the spinor summation formula is required
         \begin{equation}
 \begin{aligned}
\sum_{s}u(p,s)\bar{u}(p,s) & =\slashed{p}+m\,,\\
\sum_{s}u_{\mu}(p,s)\bar{u}_{\nu}(p,s) & =(\slashed{p}+m) \left\{ -g_{\mu\nu}+\frac{\gamma_{\mu}\gamma_{\nu}}{3}+\frac{2p_{\mu}p_{\nu}}{3m^{2}}-\frac{p_{\mu}\gamma_{\nu}-p_{\nu}\gamma_{\mu}}{3m} \right\}\,, 
    \end{aligned}
\end{equation}
for spin $\frac{1}{2}$ and $\frac{3}{2}$ respectively, and the polarization summation for massive vector meson is
        \begin{equation}
 \begin{aligned}
\sum_{\lambda_{1}}\varepsilon^{*\rho}(p_{1},\lambda_{1})\varepsilon^{\nu}(p_{1},\lambda_{1})&=-g^{\rho\nu}+\frac{p_1^{\rho}p_1^{\nu}}{m^{2}_{1}},
    \end{aligned}
\end{equation}

\section{Full expressions of amplitudes}\label{app.C}
Here, we give the full amplitudes of five $\Lambda^{0}_{b}$ decay channels we consider in this work:
         \begin{equation}
 \begin{aligned}
  \mathcal{A}(\Lambda_b^0 \rightarrow pK^-)&=\mathcal{S}(\Lambda_b^0 \rightarrow pK^-)+\mathcal{M}(D^{*-}_s,\Lambda_c^+;\bar{D}^0)+\mathcal{M}(D_s^{*-},\Lambda_c^+;\bar{D}^{*0})+\mathcal{M}(D_s^-,\Lambda_c^+;\bar{D}^{*0})+\mathcal{M}(K^-,p;\rho^0)\\
  &+\mathcal{M}(K^{*-},p;\rho^0)+\mathcal{M}(K^{*-},p;\pi^0)+\mathcal{M}(K^{*-},p;\eta)+\mathcal{M}(K^-,p;\omega)+\mathcal{M}(K^{*-},p;\omega)\\
  &+\mathcal{M}(\pi^0,\Lambda^0;K^{*+})+\mathcal{M}(\rho^0,\Lambda^0;K^{*+})+\mathcal{M}(\eta,\Lambda^0;K^{*+})+\mathcal{M}(\omega,\Lambda^0;K^{*+})+\mathcal{M}(\phi,\Lambda^0;K^{*+})\\
  &+\mathcal{M}(\bar{K}^{*0},n;\pi^+)+\mathcal{M}(\bar{K}^{*0},n;\rho^+)+\mathcal{M}(\bar{K}^0,n;\rho^+)+\mathcal{M}(\pi^0,\Lambda^0;p)+\mathcal{M}(\eta,\Lambda^0;p)+\mathcal{M}(\rho^0,\Lambda^0;p)\\
  &+\mathcal{M}(\omega,\Lambda^0;p)+\mathcal{M}(\rho^0,\Lambda^0;K^+)+\mathcal{M}(\omega,\Lambda^0;K^+)+\mathcal{M}(\phi,\Lambda^0;K^+)
    \end{aligned}
\end{equation}

         \begin{equation}
 \begin{aligned}
\mathcal{A}(\Lambda_b^0\rightarrow p\pi^-) &=\mathcal{S}(\Lambda_b^0\rightarrow p\pi^-)+\mathcal{M}(D^-,\Lambda_c^+;\bar{D}^{*0})+\mathcal{M}(D^{*-},\Lambda_c^+;\bar{D}^{*0})+\mathcal{M}(D^{*-},\Lambda_c^+;\bar{D}^0)+\mathcal{M}(D^-,\Lambda_c^+;\Sigma_c^{++})\\
    &+\mathcal{M}(D^{*-},\Lambda_c^+;\Sigma_c^{++})+\mathcal{M}(\pi^-,p;\rho^0)+\mathcal{M}(\rho^-,p;\pi^0)+\mathcal{M}(\rho^-,p;\omega)+\mathcal{M}(\pi^-,p;\Delta^{++})\\
    &+\mathcal{M}(\rho^-,p;\Delta^{++})+\mathcal{M}(\pi^0,n;\rho^+)+\mathcal{M}(\rho^0,n;\pi^+)+\mathcal{M}(\omega,n;\rho^+)+\mathcal{M}(\pi^0,n;p)\mathcal{M}(\eta,n;p)\\
    &+\mathcal{M}(\rho^0,n;p)+\mathcal{M}(\omega,n;p)+\mathcal{M}(K^0,\Lambda^0;K^{*+})+\mathcal{M}(K^{*0},\Lambda^0;K^+)+\mathcal{M}(K^{*0},\Lambda^0;K^{*+})\\
    &+\mathcal{M}(K^0,\Lambda^0;\Sigma^+)+\mathcal{M}(K^{*0},\Lambda^0;\Sigma^+)+\mathcal{M}(K^0,\Lambda^0;\Sigma^{*+})+\mathcal{M}(K^{*0},\Lambda^0;\Sigma^{*+})+\mathcal{M}(\pi^0,n;\Delta^+)\\
    &+\mathcal{M}(\rho^0,n;\Delta^+)
   \end{aligned}
\end{equation}

         \begin{equation}
 \begin{aligned}
\mathcal{A}(\Lambda_b^0\rightarrow pK^{*-})&=\mathcal{S}(\Lambda_b^0\rightarrow pK^{*-})+\mathcal{M}(D_s^-,\Lambda_c^+;\bar{D}^0)+\mathcal{M}(D_s^-,\Lambda_c^+;\bar{D}^{*0})+\mathcal{M}(D_s^{*-},\Lambda_c^+;\bar{D}^0)+\mathcal{M}(D_s^{*-},\Lambda_c^+;\bar{D}^{*0})\\
        &+\mathcal{M}(K^-,p;\pi^0)+\mathcal{M}(K^-,p;\rho^0)+\mathcal{M}(K^-,p;\eta)+\mathcal{M}(K^{*-},p;\pi^0)+\mathcal{M}(K^{*-},p;\eta)+\mathcal{M}(K^{*-},p;\rho^0)\\
        &+\mathcal{M}(K^{*-},p;\omega)+\mathcal{M}(\eta,\Lambda^0;K^+)+\mathcal{M}(\eta,\Lambda^0;K^{*+})+\mathcal{M}(\pi^0,\Lambda^0;K^+)+\mathcal{M}(\pi^0,\Lambda^0;K^{*+})+\mathcal{M}(\rho^0,\Lambda^0;K^+)\\
        &+\mathcal{M}(\rho^0,\Lambda^0;K^{*+})+\mathcal{M}(\phi,\Lambda^0;K^+)+\mathcal{M}(\phi,\Lambda^0;K^{*+})+\mathcal{M}(\omega,\Lambda^0;K^+)+\mathcal{M}(\omega,\Lambda^0;K^{*+})+\mathcal{M}(\eta,\Lambda^0;p)\\
        &+\mathcal{M}(\pi^0,\Lambda^0;p)+\mathcal{M}(\rho^0,\Lambda^0;p)+\mathcal{M}(\omega,\Lambda^0;p)+\mathcal{M}(\bar{K}^0,n;\pi^+)+\mathcal{M}(\bar{K}^0,n;\rho^+)+\mathcal{M}(\bar{K}^{*0},n;\pi^+)\\
        &+\mathcal{M}(\bar{K}^{*0},n;\rho^+)
   \end{aligned}
\end{equation}

         \begin{equation}
 \begin{aligned}
\mathcal{A}(\Lambda_b^0\rightarrow p\rho^-)&=\mathcal{S}(\Lambda_b^0 \rightarrow p\rho^-)+\mathcal{M}(D^-,\Lambda_c^+;\bar{D}^0)+\mathcal{M}(D^-,\Lambda_c^+;\bar{D}^{*0})+\mathcal{M}(D^{*-},\Lambda_c^+;\bar{D}^0)+\mathcal{M}(D^{*-},\Lambda_c^+;\bar{D}^{*0})\\
    &+\mathcal{M}(D^-,\Lambda_c^+;\Sigma_c^{++})+\mathcal{M}(D^{*-},\Lambda_c^+;\Sigma_c^{++})+\mathcal{M}(\pi^-,p;\pi^0)+\mathcal{M}(\pi^-,p;\omega)+\mathcal{M}(\rho^-,p;\eta
    )\\
    &+\mathcal{M}(\rho^-,p;\rho^0)+\mathcal{M}(\pi^-,p;\Delta^{++})+\mathcal{M}(\rho^-,p;\Delta^{++})+\mathcal{M}(\pi^0,n;\pi^+)+\mathcal{M}(\pi^0,n;p)+\mathcal{M}(\pi^0,n;\Delta^+)\\
    &+\mathcal{M}(\eta,n;\rho^+)+\mathcal{M}(\eta,n;p)+\mathcal{M}(\rho^0,n;\rho^+)+\mathcal{M}(\rho^0,n;p)+\mathcal{M}(\rho^0,n;\Delta^+)+\mathcal{M}(\omega,n;\pi^+)\\
    &+\mathcal{M}(\omega,n;p)+\mathcal{M}(K^0,\Lambda^0;K^+)+\mathcal{M}(K^0,\Lambda^0;K^{*+})+\mathcal{M}(K^0,\Lambda^0;\Sigma^+)+\mathcal{M}(K^0,\Lambda^0;\Sigma^{*+})\\
    &+\mathcal{M}(K^{*0},\Lambda^0;K^+)+\mathcal{M}(K^{*0},\Lambda^0;K^{*+})+\mathcal{M}(K^{*0},\Lambda^0;\Sigma^+)+\mathcal{M}(K^{*0},\Lambda^0;\Sigma^{*+})
   \end{aligned}
\end{equation}

         \begin{equation}
 \begin{aligned}
\mathcal{A}(\Lambda_b^0\rightarrow \Lambda^0 \phi)&=\mathcal{S}(\Lambda_b^0\rightarrow \Lambda^0\phi)+\mathcal{M}(D^{-}_s,\Lambda_c^+;D^{-}_s)+\mathcal{M}(D^{*-}_s,\Lambda_c^+;D^{*-}_s)+\mathcal{M}(D^{*-}_s,\Lambda_c^+;D^{-}_s)+\mathcal{M}(D^{-}_s,\Lambda_c^+;D^{*-}_s)\\
    &+\mathcal{M}(K^-,p;K^-)+\mathcal{M}(K^-,p;K^{*-})+\mathcal{M}(K^{*-},p;K^-)+\mathcal{M}(K^{*-},p;K^{*-})+\mathcal{M}(\eta,\Lambda^0;\phi)\\
    &+\mathcal{M}(\phi,\Lambda^0;\eta)+\mathcal{M}(\bar{K}^0,n;\bar{K}^0)+\mathcal{M}(\bar{K}^0,n;\bar{K}^{*0})+\mathcal{M}(\bar{K}^{*0},n;\bar{K}^0)+\mathcal{M}(\bar{K}^{*0},n;\bar{K}^{*0})+\mathcal{M}(\eta,\Lambda^0;\Lambda^0)\\
    &+\mathcal{M}(\omega,\Lambda^0;\Lambda^0)+\mathcal{M}(\phi,\Lambda^0;\Lambda^0)
   \end{aligned}
\end{equation}

\end{document}